\documentclass[a4paper,11pt]{article}
\usepackage{jheppub} 
\usepackage{lineno}
\usepackage{rotating}
\usepackage{array}
\usepackage{amsmath}
\usepackage{bbold}
\usepackage[normalem]{ulem}
\usepackage{slashed}
\usepackage{booktabs}
\usepackage[pdftex,table,dvipsnames]{xcolor}
\usepackage{units}
\usepackage{multirow}
\usepackage{url}
\usepackage{parskip}
\usepackage[all]{nowidow}
\usepackage{placeins}

\usepackage{longtable}

\usepackage{xspace}
\usepackage{subfig}
\usepackage{color}

\usepackage{dsfont}
\usepackage{soul}
\usepackage{hyperref}
\usepackage{physics}
\usepackage{cleveref}

\usepackage{physics} 
\usepackage{bm} 

\newcommand{\bc}{\bm{c}}

\newcommand{\nn}{\notag\\}
\newcommand{\muCPT}{\mu_{\mathrm{chPT}}}
\newcommand{\cpt}{\mathrm{chPT}}
\newcommand{\NR}{\mathrm{NR}}
\newcommand{\MeV}{\mathrm{MeV}}
\newcommand{\GeV}{\mathrm{GeV}}
\newcommand{\TeV}{\mathrm{TeV}}
\newcommand{\mC}{\mathcal{C}}
\newcommand{\mL}{\mathcal{L}}
\newcommand{\mM}{\mathcal{M}}
\newcommand{\mN}{\mathcal{N}}
\newcommand{\mO}{\mathcal{O}}
\newcommand{\bmc}{\bm{c}}
\newcommand{\bD}{\bm{D}}

\newcommand{\bl}{\bm{l}}
\newcommand{\bmm}{\bm{m}}
\newcommand{\bq}{\bm{q}}
\newcommand{\br}{\bm{r}}
\newcommand{\bu}{\bm{u}}
\newcommand{\bv}{\bm{v}}
\newcommand{\bGamma}{\bm{\Gamma}}

\newcommand{\bxi}{\bm{\xi}}
\newcommand{\bsigma}{\bm{\sigma}}
\newcommand{\bSigma}{\bm{\Sigma}}
\newcommand{\absq}{\abs{\bm{q}}}

\preprint{Nikhef 2025-017}

\title{\boldmath ALP-mediated Dark Matter-Nucleon Scattering}

\author[1,2]{Wim Beenakker,}
\author[1]{Dani\"el Mikkers,}
\author[1,2]{Anh Vu Phan,}
\author[1,2]{Susanne Westhoff\,}

\affiliation[1]{Institute for Mathematics, Astrophysics and Particle Physics, Radboud University, 6500 GL \mbox{Nijmegen}, The Netherlands}
\affiliation[2]{Nikhef, Science Park 105, 1098 XG Amsterdam, The Netherlands}

\emailAdd{w.beenakker@science.ru.nl}
\emailAdd{danielmikkers@proton.me}
\emailAdd{anhvu.phan@ru.nl}
\emailAdd{susanne.westhoff@ru.nl}

\abstract{We perform a comprehensive analysis of dark matter-nucleon scattering via the exchange of axion-like particles (ALPs). At first sight, this might appear of little practical use, as non-relativistic scattering through pseudo-scalar interactions is momentum-suppressed and spin-dependent, resulting in scattering rates below any experimental sensitivity. We show that the scattering rate can be drastically enhanced in two ways. First, light ALPs with masses below the typical momentum transfer at direct detection experiments lift the momentum suppression by acting as essentially massless mediators. Second, ALP exchange through loops induces coherent spin-independent scattering. If the ALP has flavor-changing couplings to up-type quarks, loop-induced scattering receives an extra strong enhancement by the top-quark mass. We deduce that, contrary to common lore, XENONnT and PandaX-4T are already sensitive to ALP-mediated dark matter-nucleon scattering. The next generation of direct detection experiments will probe far into the parameter space of the ALP effective theory, potentially exceeding the sensitivity of collider searches.}

\begin{document}
\maketitle
\flushbottom

\section{Introduction}
\label{sec:intro}
Over the past years, dark matter direct detection experiments have reached a remarkable sensitivity to dark matter-nucleon scattering~\cite{Schumann:2019eaa,Billard:2021uyg,Akerib:2022ort}. In many models for particle dark matter (DM), the null results of searches set the strongest constraints on the underlying fundamental interactions with quarks and gluons~\cite{Cirelli:2024ssz,Arcadi:2024ukq,Bernal:2017kxu}. Classified by their Lorentz structure, scalar and vector interactions induce coherent spin-independent scattering. Pseudo-scalar and axial-vector interactions lead to spin-dependent scattering, which is comparatively suppressed for heavy nuclei~\cite{Goodman:1984dc,Agrawal:2010fh}. In addition, pseudo-scalar interactions are momentum-suppressed in the non-relativistic regime probed by experiments. For Xenon nuclei, scattering through pseudo-scalar interactions is about 8 orders of magnitude smaller than
 scattering through scalar or vector interactions.

It seems a logical consequence that dark matter-nucleon scattering through pseudo-scalar mediators has been underexplored. Among the one-particle portals to a dark sector~\cite{Pospelov:2007mp,Jaeckel:2010ni,Chu:2011be}, Higgs-mixing scalars~\cite{Djouadi:2012zc,Freitas:2015hsa,Arcadi:2019lka} and dark photons~\cite{Essig:2011nj,Fabbrichesi:2020wbt} have been established as benchmark scenarios to compare the sensitivity of direct detection experiments among themselves and against other dark matter searches. On the contrary, axion-like particles (ALPs) as pseudo-scalar mediators of dark matter-nucleon scattering have received much less attention. This lack of exploration makes it difficult, if not impossible, to estimate the detection prospects of ALP-mediated dark matter and link the corresponding searches to other experimental probes.

In this work, we perform the first comprehensive analysis of dark matter-nucleon scattering through ALPs. We show that ALP-mediated scattering is actually observable at current direct detection experiments, provided that the ALP is light compared to the momentum exchange with nucleons or has flavor-changing couplings to top quarks.

Let us briefly review the state of the art in dark matter-nucleon scattering through pseudo-scalar mediators in general and ALPs in particular. Dark matter models which induce primarily spin-dependent interactions have been classified in~\cite{Agrawal:2010fh}. For generic pseudo-scalars, it was found that the momentum suppression of spin-dependent scattering can be lifted by a light mediator~\cite{Freytsis:2010ne} and that many models generate spin-independent scattering at the loop level~\cite{Freytsis:2010ne,Ipek:2014gua,Arcadi:2017wqi}. The lifting of momentum suppression for light ALPs has also been investigated  in dark matter-electron scattering~\cite{Buttazzo:2020vfs}.

If the pseudo-scalar mediator is an ALP, the phenomenology changes because ALP interactions respect a shift symmetry, which prevents fundamental couplings of ALPs with particles of the Standard Model (SM). ALP interactions are therefore described by an effective field theory (EFT) with a cutoff scale far above the momentum scale of dark matter-nucleon scattering~\cite{Georgi:1986df}. As a consequence, the suppression of spin-dependent scattering at tree level over spin-independent scattering at loop level is reduced: The former is suppressed quadratically, the latter by four powers of the cutoff scale. This viewpoint is reflected in recent analyses of ALP-mediated dark matter, which claim dark matter-nucleon scattering to be suppressed below the reach of current experiments~\cite{Fitzpatrick:2023xks,Armando:2023zwz}. As we will show, this conclusion is circumvented in the presence of flavor-changing ALP couplings, which raise spin-independent scattering to an observable level through a new enhancement mechanism.

ALPs are not only viable mediators of a dark force today, but also throughout the cosmic history. For fermion dark matter, the cosmologically relevant parameter space has been mapped out for generic ALP mediators with couplings to gluons~\cite{Fitzpatrick:2023xks} and leptons~\cite{Armando:2023zwz}, and for the QCD axion as a mediator~\cite{Dror:2023fyd}. It was found that the observed relic dark matter abundance can be obtained from various cosmic histories, including freeze-in and freeze-out scenarios. These cosmological benchmarks can give valuable hints for ALP-mediated dark matter searches at laboratory experiments.

Our analysis of ALP-mediated dark matter-nucleon scattering is organized as follows. In~\cref{sec:model}, we review the ALP effective theory and derive experimental bounds on the relevant ALP couplings with quarks and gluons. In~\cref{sec:dark matter-nucleus}, we discuss the formalism for non-relativistic dark matter-nucleus scattering through contact interactions and through light mediators within the ALP effective theory. Section~\ref{sec:tree} is devoted to spin-dependent scattering through heavy and light ALPs at tree level. In~\cref{sec:loop}, we calculate loop-induced spin-independent scattering for three separate contributions: flavor-diagonal ALP couplings in~\cref{sec:flavor-diagonal-loop}, flavor-changing ALP couplings in~\cref{sec:flavor-changing-loop}, and non-perturbative contributions from light meson exchange in chiral perturbation theory (chPT) in~\cref{sec:chiral-loop}. We also comment on potential contributions from higher-order couplings in the ALP effective theory in~\cref{sec:ho}. In~\cref{sec:results}, we quantify all these individual contributions to ALP-mediated dark matter-nucleon scattering and provide numerical predictions of event rates at current and future direct detection experiments. We conclude in~\cref{sec:conclusion}. Technical details on the loop functions and the chiral Lagrangian are relegated to~\cref{app:analytics}.

\section{ALP effective theory and dark matter}
\label{sec:model}
In this section, we briefly review the effective theory of ALP interactions and introduce the relevant couplings to SM particles and dark matter. Throughout this work, we assume that dark matter consists of Dirac fermions. Other dark matter candidates are possible, but would lead to a different phenomenology of dark matter-nucleon scattering. They deserve a separate analysis.

ALP couplings are constrained by a number of searches at collider experiments and by astrophysical observables. For the relevant couplings in dark matter-nucleon scattering, we derive the current bounds from colliders over a broad range of ALP masses. Astrophysics typically provides stronger bounds than colliders on ALPs with masses below about 10\,MeV, see e.g.~\cite{Raffelt:2006cw,Depta:2020wmr,Caputo:2024oqc}. At this mass scale, the ALP is essentially massless in dark matter-nucleon scattering at current experiments and the scattering phenomenology is similar to somewhat heavier ALPs, which do not leave a trace in astrophysics. We therefore do not consider astrophysical bounds in this work.

\subsection{ALP couplings to Standard Model particles and fermion dark matter}
The ALP effective theory includes all ALP interactions with SM particles that preserve a shift symmetry of the ALP field, $a\to a + c$, with $c$ being a constant. We assume that this shift symmetry is also preserved by the ALP coupling to dark matter fermions. In this case, the effective Lagrangian for ALP interactions that are relevant for dark matter-nucleon scattering reads~\cite{Georgi:1986df,Bauer:2020jbp}\footnote{The case of pseudoscalar-mediated dark matter interactions without a shift symmetry has been considered in~\cite{Alanne:2017oqj}.}
\begin{align}\label{eq:lagrangian}
    \mathcal{L}_{\rm eff}(\mu) & = \frac{1}{2}\,\partial_\mu a\,\partial^\mu a - \frac{m_a^2}{2}\,a^2 + i\bar\chi \gamma^\mu\partial_\mu \chi - m_\chi \bar\chi \chi + \frac{c_\chi}{2}\frac{\partial^\mu a}{f_a} \bar \chi\, \gamma_\mu\gamma_5 \chi\\\nonumber
    & \quad + \frac{\partial^\mu a}{f_a}\, \bar U_L\, \bmc_U \gamma_\mu U_L + \frac{\partial^\mu a}{f_a}\, \bar D_L \bmc_D \gamma_\mu D_L + \frac{\partial^\mu a}{f_a}\, \bar U_R\, \bmc_u \gamma_\mu U_R + \frac{\partial^\mu a}{f_a}\, \bar D_R\, \bmc_d \gamma_\mu D_R \\\nonumber
    & \quad  + c_{GG}\,\frac{a}{f_a}\,\frac{\alpha_s}{4\pi}\,G_{\mu\nu}^b \widetilde{G}^{b,\mu\nu}\,.
\end{align}
Here $a$ is the ALP field with associated mass $m_a$, and $\chi$ is a Dirac fermion without any charge under the SM gauge group. Further, $G_{\mu\nu}^b$ denotes the gluon field strength tensor and its dual is $\widetilde{G}^{b,\mu\nu} = \frac{1}{2}\,\epsilon^{\mu\nu\rho\sigma}G_{\rho\sigma}^b$, with $SU(3)_C$ gauge indices $b = \{1,\dots 8\}$ and the strong coupling $\alpha_s$. The quark fields correspond to mass eigenstates; $U,\,D$ are 3-vectors and $\bmc_i$ are $3\times 3$ matrices, both in flavor space. The ALP couplings to left-handed up- and down-type quarks are related by the CKM matrix $V$ via $\bmc_D = V^\dagger \bmc_U V$. All parameters in this Lagrangian are defined at a scale $\mu < \Lambda = 4\pi f_a$ below the cutoff scale $\Lambda$. Throughout this work, we set $f_a = 1\,$TeV, reflecting the fact that so far no fundamental particles beyond the Standard Model have been observed below the TeV scale.

It is convenient to write the ALP couplings to up-type quarks in terms of axial-vector and vector parts,
\begin{align}\label{eq:ALP-quark-couplings}
    \mathcal{L}_{\rm eff} \supset \frac{\partial^\mu a}{2f_a}\, \bar U\, (\bmc_u - \bmc_U) \gamma_\mu \gamma_5\, U + \frac{\partial^\mu a}{2f_a}\, \bar U \, (\bmc_u + \bmc_U) \gamma_\mu\, U,
\end{align}
and analogously for down-type quarks with $u\to d$ and $U \to D$. For on-shell quarks, applying integration by parts and the Dirac equation to~\eqref{eq:ALP-quark-couplings} leads to
\begin{align}
    \mathcal{L}_{\rm eff} \supset -\frac{i a}{2 f_a} \sum_{i,j}\left[ (m_{u_i} + m_{u_j})\,\bar U_i (\bmc_u - \bmc_U)_{ij}\gamma_5 U_j + (m_{u_i} - m_{u_j})\,\bar U_i (\bmc_u + \bmc_U)_{ij} U_j \right], \label{eq:flavor-structure}
\end{align}
with the generation indices $i,j=\{1,2,3\}$. The flavor-diagonal (FD) ALP couplings to up-type quarks,
\begin{align}
c_{uu} = (\bmc_u - \bmc_U)_{11}\,,\quad c_{cc} = (\bmc_u - \bmc_U)_{22}\,,\quad c_{tt} = (\bmc_u - \bmc_U)_{33}\,,
\end{align}
originate from the axial-vector current in~\eqref{eq:ALP-quark-couplings}. Analogous expressions hold for down-type quarks. The vector part $(\bmc_u + \bmc_U)_{ii}$ is unobservable in processes which preserve flavor-diagonal vector currents, including QED, QCD and dark-sector interactions.
Flavor-changing (FC) couplings, in turn, receive both axial-vector and vector contributions. In this work, the relevant flavor-changing couplings are the up-top couplings
\begin{align}\label{eq:FC-couplings}
    c_{ut}^A = (\bmc_u - \bmc_U)_{13}\,,\quad c_{ut}^V = (\bmc_u + \bmc_U)_{13}
\end{align}
and couplings in the down-quark sector
\begin{align}\label{eq:FC-couplings-down}
   c_{d_i d_j}^A = (\bmc_d - \bmc_D)_{ij}\,,\quad c_{d_i d_j}^V = (\bmc_d + \bmc_D)_{ij}\,,
\end{align}
where the quark flavors are denoted as $d_1 = d$, $d_2 = s$, $d_3 = b$. Hermiticity of the Lagrangian requires $c_{ij}^{A,V} = (c_{ji}^{A,V})^\ast$.

\subsection{Experimental bounds on ALP couplings to light quarks and gluons}\label{sec:fd-bounds}
The couplings of ALPs to particles of the Standard Model are experimentally constrained by a large number of measurements and searches at particle physics experiments. For this work, ALP couplings to quarks and gluons are most relevant. In this section, we summarize existing bounds on the ALP-gluon coupling and on flavor-diagonal ALP couplings to up and down quarks.

When deriving bounds on the various ALP couplings, we assume that only the considered coupling is present at the cutoff scale $\Lambda$. Additional couplings can change the derived bounds in the observables we consider and make other observables sensitive to the coupling of interest.

The ALP coupling to dark matter is not constrained independently of the ALP-SM couplings by collider searches. We therefore apply only the loose perturbativity bound $c_{\chi} < 4\pi$. We also neglect potential ALP decays to dark matter, $a\to \chi\bar\chi$. If present, they would strengthen the bounds on other ALP couplings from searches for $K\to \pi a$ and $B\to (K,\pi)\,a$ decays with invisible ALPs.

\paragraph{ALP decays}
ALPs with couplings to gluons and quarks of the first generation decay mostly hadronically, provided that the ALP mass lies above the three-pion threshold, $m_a > 3 m_\pi$. For ALP masses in the perturbative regime, $m_a \gg 2\,\GeV$, the decay width for ALPs into light-flavored hadrons is given by~\cite{Bauer:2017ris,Bauer:2021mvw}
\begin{align}
    \Gamma(a\to \text{hadrons}) \approx \frac{\abs{C^{\rm eff}_{GG} (m_a)}^2}{23\,\rm{nm}} \qty(\frac{\alpha_s(m_a)}{0.3})^2 \qty(\frac{m_a}{2\,\GeV})^3 \qty(\frac{1\,\TeV}{f_a})^2\,.
\end{align}
We quote the decay width in units of inverse length to give an impression of the corresponding proper decay length. For $m_a \ll m_t$, the effective ALP-gluon coupling can be expressed in terms of the couplings at the cutoff scale as
\begin{align}
    C^{\rm eff}_{GG}(m_a) \approx 0.96\,c_{GG} + 0.48\, \qty[c_{uu}(\Lambda) + c_{dd} (\Lambda)]\,.
\end{align}
For ALP masses $3 m_{\pi} < m_a < 2\,\GeV$, hadronic and semi-hadronic decays such as $a\to 3\pi$, $a\to \pi\pi\gamma$, $a\to \pi\pi\eta$ determine the decay width~\cite{Ovchynnikov:2025gpx,Balkin:2025enj}.
For $3 m_\pi \ll m_a \lesssim 0.6\,$GeV, the hadronic decay width is well approximated by the sum of $3\pi^0$ and $\pi^+\pi^-\pi^0$ final states,\footnote{The decay width including the full phase-space dependence can be found in~\cite{Bauer:2021mvw}.}
\begin{align}
    \Gamma(a\to3\pi) \approx \frac{\qty[0.73\, c_{GG} + c_{uu}(\Lambda) - c_{dd}(\Lambda)]^2}{1\,\rm{mm}}  \qty(\frac{m_a}{1\,\GeV})^3 \qty(\frac{1\,\TeV}{f_a})^2.
\end{align}
ALPs with masses below the hadronic threshold, $m_a < 3 m_{\pi}$, decay mostly into two photons through renormalization group (RG) effects of ALP-gluon and ALP-quark couplings present at the cutoff scale. The partial decay width and branching ratio for $a\to \gamma\gamma$ can be found in \cite{Bauer:2020jbp,Bauer:2021mvw,Balkin:2025enj}.
Depending on the ALP properties and the setup of the experiment, the ALP might decay mostly outside the detector and appear as invisible.

\paragraph{ALP production in meson decays}
Light ALPs can be produced in the meson decays $K\to \pi a$ and $B\to K a$ through the flavor-changing neutral currents $c_{ds}^V$ and $c_{sb}^V$, defined in~\cref{eq:FC-couplings-down}. We assume that $c^A_{d_id_j} (\Lambda) = c^V_{d_id_j}(\Lambda) = 0$ for $i \ne j$, which implies $\bc_U = 0$. In this case, the meson decays are loop-induced, which leads to the smallest decay widths and most conservative bounds on the ALP couplings $c_{GG}$, $c_{uu}$ and $c_{dd}$.
For kaon decays, the branching ratio into ALPs has been calculated in chPT~\cite{Bauer:2021mvw}, yielding
\begin{align}
    \mathcal{B}(K^- \to \pi^- a) &\approx 7\cdot 10^{-7} \,
    \frac{\qty[c_{GG} + 0.5\, c_{uu} (\Lambda) + 0.2\, c_{dd}(\Lambda)]^2}{(f_a/1\,\TeV)^2} \lambda^{1/2} \qty(\frac{m_\pi^2}{m_K^2},\frac{m_a^2}{m_K^2})\,,
\end{align}
with $\lambda(x,y) = 1 + x^2 + y^2 - 2(x+y+xy)$. For $B$ decays, the branching ratio into ALPs can be calculated in perturbation theory, see e.g.~\cite{Ferber:2022rsf}), which yields 
\begin{align}
    \mathcal{B}(B^-\to K^- a)\approx4.6\cdot 10^{-6}\,\frac{\qty[c_{GG} + 0.5\, c_{uu} (\Lambda) + 0.5\, c_{dd}(\Lambda)]^2}{(f_a/1\,\TeV)^2} \frac{f_0^2(m_a^2)}{f_0^2(0)} \lambda^{1/2} \qty(\frac{m_K^2}{m_B^2},\frac{m_a^2}{m_B^2}).
\end{align}
Here we have used the scalar hadronic form factor $f_0(0)$ from~\cite{Gubernari:2018wyi}.

\paragraph{ALP couplings to gluons}
At the NA62 experiment, ALPs can be produced from kaon decays $K\to \pi a$. In this mass region, hadronic decays such as $a\to 3\pi$ are kinematically forbidden and the dominant decay mode is $a \to \gamma\gamma$. For ALPs with masses 
$m_a \lesssim \text{min}(0.126\, |c_{GG}|^{-2/3},0.6)\,$GeV, assuming hadronic decays $a\to 3\pi$ only and $f_a = 1\,$TeV, the proper lifetime exceeds $5\,$ns, except for ALP masses close to the pion mass. For $m_a < m_K - m_\pi$, this condition is fulfilled for $|c_{GG}| < 0.2$. In this case, the produced ALPs decay mostly outside the NA62 detector and the search for $K^+ \to \pi^+ X$ with an invisible particle $X$~\cite{NA62:2025upx} applies. The sensitivity of this search to the ALP-gluon coupling depends on the ALP mass. For $m_a > m_\pi$, NA62 finds an upper bound of $\mathcal{B}(K^+ \to \pi^+ X) \lesssim 10^{-11}$, which corresponds to
\begin{align}\label{eq:cgg-bound-na62}
m_\pi < m_a < m_K-m_\pi: \quad \frac{|c_{GG}|}{f_a} \lesssim \frac{0.0038}{\text{TeV}}\,.
\end{align}
For $m_a < m_\pi$, the bound on the branching ratio is a factor of $2-3$ weaker. Notice that the search excludes the mass region of strong ALP-pion mixing, $m_a \approx m_\pi$, which is constrained by a separate search for $K^+\to \pi^+\pi^0,\pi^0\to\ $\emph{invisible} by NA62~\cite{NA62:2020pwi}.

For couplings $|c_{GG}| > 0.2$, the ALP decay length is reduced and the search loses sensitivity. In this parameter region, searches for $K_L\to \pi^0\nu\bar\nu$~\cite{Bauer:2021mvw} and $B^+\to K^+\nu\bar \nu$~\cite{Abumusabh:2025zsr,Ferber:2022rsf} exclude all of the ALP parameter space for $m_a < 100\,$MeV. Heavier ALPs with strong gluon couplings are constrained by searches for di-muon and di-photon final states~\cite{Bauer:2021mvw}. The strongest bound in this mass region is due to a search for $B^\pm \to K^\pm a,a\to\gamma\gamma$ with proper decay length $c\tau_a \lesssim 1\,$mm at the BaBar experiment~\cite{BaBar:2021ich}.
Using the experimental limit $\mathcal{B}(B^-\to K^- a) \mathcal{B}(a \to \gamma\gamma) \lesssim 2\times 10^{-7}$ for $m_a < 1$ GeV, we have
\begin{align}
    0.175\,\GeV< m_a \lesssim 1\,\GeV&:\quad \frac{|c_{GG}|}{f_a} \lesssim \frac{2.1}{\text{TeV}} \qty[\frac{\mathcal{B}(a \to \gamma\gamma)}{10^{-2}}]^{-1/2}\,.
\end{align}    
For $\mathcal{B}(a \to \gamma\gamma) > 0.1$, valid for most ALP masses $m_a < 0.9\,$GeV, this excludes ALP couplings $|c_{GG}|/f_a > 0.2/$TeV.
For $m_a \gtrsim 1\,\GeV$, the branching fraction into di-photons is strongly suppressed due to the overwhelming decay into light-flavored hadrons.

For ALPs too heavy to be resonantly produced in meson decays, dijet production at the LHC offers a direct probe of the ALP-gluon coupling. In particular, angular correlations in dijet distributions are modified by the exchange of a virtual ALP, leading to the loose bound~\cite{Gavela:2019cmq,Bruggisser:2023npd}
\begin{align}
m_a \ll M_{jj}: \quad \frac{|c_{GG}|}{f_a} \lesssim \frac{100}{\text{TeV}}\,.
\end{align}
For ALP masses well below the dijet invariant mass, $M_{jj}$, this bound is largely independent of the ALP mass.

\paragraph{Flavor-diagonal ALP couplings to up and down quarks}
As described above, ALPs with couplings to up and down quarks have a similar phenomenology as ALPs with gluon couplings. With pure ALP couplings to light quarks, ALPs below the hadronic threshold decay dominantly via $a \to \gamma\gamma$, with a proper lifetime of $\tau_a \gtrsim 5\,$ns for $m_a \lesssim 0.270\,\abs{c_{uu}(\Lambda) - c_{dd}(\Lambda)}^{-2/3}\GeV$, assuming $f_a = 1\,$TeV.
 Using NA62's search for $K^+\to\pi^+ X$ again~\cite{NA62:2025upx}, we obtain
\begin{align}\label{eq:cuu-bound-na62}
m_\pi < m_a < m_K-m_\pi : \quad \frac{|c_{uu}(\Lambda) + 0.4\,c_{dd}(\Lambda)|}{f_a} \lesssim \frac{0.0076}{\text{TeV}}\,,
\end{align}
and somewhat looser bounds for ALP masses below the pion mass.
These bounds apply for ALP couplings $\abs{c_{uu}(\Lambda) - c_{dd}(\Lambda)} < 0.67$, due to the lifetime sensitivity. The parameter space of larger couplings is excluded by searches for
$K_L\to \pi^0\nu\bar\nu$ and $B^+\to K^+\nu\bar \nu$, as for ALP-gluon couplings.

For ALPs heavier than kaons, hadronic decays dominate and the ALP is short-lived at detector scales. The currently strongest bound is due to the same BaBar search for $B^+\to K^+a,a\to\gamma\gamma$ mentioned earlier~\cite{BaBar:2021ich}.
 Using the limit $\mathcal{B}(B^-\to K^- a) \mathcal{B}(a \to \gamma\gamma) \lesssim 2\times 10^{-7}$ for $m_a < 1$ GeV again, we deduce
\begin{align}
    0.175\,\GeV< m_a \lesssim 1\,\GeV&:\quad \frac{|c_{uu}(\Lambda) + c_{dd}(\Lambda)|}{f_a}\lesssim \frac{4.2}{\text{TeV}} \qty[\frac{\mathcal{B}(a \to \gamma\gamma)}{10^{-2}}]^{-1/2}\,.
\end{align} 
For ALPs with $m_a \gtrsim 1\,\GeV$, the ALP decays overwhelmingly into final states with light hadrons such as $\eta\pi\pi$~\cite{Balkin:2025enj}. As no searches for these final states exist, we lack a bound on $c_{uu}$ and $c_{dd}$ in this ALP mass region. In the perturbative QCD regime with $m_a \gg 2\,$GeV, ALP couplings to light quarks are strongly suppressed with the quark mass. Collider searches therefore do not impose any bounds on $c_{uu}$ and $c_{dd}$ in this mass region due to small production rates.

\subsection{Experimental bounds on flavor-changing ALP couplings}\label{sec:fc-bounds}
In ALP-mediated dark matter-nucleon scattering, the flavor-changing ALP couplings $c_{ut}^{V,A}$ from~\cref{eq:FC-couplings} play a special role for reasons that will become clear later. In this section, we derive bounds on $c_{ut}^{V,A}$ from collider searches for ALPs in a broad range of masses. As before, we assume that only this coupling is present at the cutoff scale of the ALP effective theory, and we neglect ALP decays to dark matter.

\paragraph{ALP decays} ALPs with masses larger than the top mass, $m_a > m_t$, decay mostly through $a \to t\bar u$. For real couplings $c_{ut}^{V,A}$, the corresponding decay width reads
\begin{align}
    \Gamma_{a \to t\bar u} = \Gamma_{a \to \bar tu} = \frac{3 m_a}{32\pi} \qty[(c^V_{ut})^2+(c^A_{ut})^2] \frac{m_t^2}{f_a^2} \qty(1 - \frac{m_t^2}{m_a^2})^2.
\end{align}
Such heavy ALPs decay promptly within the detector, unless the couplings are strongly suppressed.

Through the CKM relation $\bc_D = V^\dagger \bc_U V$, the ALP couplings $c_{ut}^{V,A}$ induce flavor-changing down-quark couplings at tree level~\cite{MartinCamalich:2020dfe,Bauer:2020jbp},
\begin{align}\label{eq:cD-ij}
    (\bc_D)_{ij} = \frac12\qty[ V^*_{1i} (c^V_{ut} - c^A_{ut}) V_{3j} + V^*_{3i} (c^V_{tu} - c^A_{tu})V_{1j}] \stackrel{\rm real}{=} \frac{1}{2}(c^V_{ut} - c^A_{ut}) ( V^*_{1i}  V_{3j} + V^*_{3i} V_{1j})\,.
\end{align}
Throughout our analysis, we will assume real ALP couplings $c_{ut}^{V,A}$, so that the last relation applies. These couplings induce hadronic ALP decays, which dominate for $m_a < m_t$ wherever kinematically allowed.

For ALPs with masses $\muCPT < m_a < m_t$, these hadronic decays can be calculated perturbatively. In the limit $m_a \gg m_{d_i},m_{d_j}$, the partial decay width reads
\begin{align}
    \Gamma_{a\to d_i \bar d_j} = \frac{3 m_a}{16\pi} \,|(\bc_D)_{ij}|^2 \frac{m_{d_i}^2 + m_{d_j}^2}{f_a^2}.
\end{align} 
Due to the CKM structure of the coupling and the 
quark mass dependence of the decay rate, the ALP decays dominantly into a $b\bar d$ or $\bar b d$ pair, if kinematically allowed. The corresponding decay length is
\begin{align}
    \lambda_a \approx \frac{26.4\,\rm pm}{|c^V_{ut} - c^A_{ut}|^2} \frac{10\,\GeV}{m_a}\qty(\frac{5\,\GeV}{m_b})^2 \qty(\frac{f_a}{1\,\rm TeV})^2. \label{eq:ALP-decaylength-above5GeV}
\end{align}
ALPs with $m_a > m_B$ decay promptly at colliders, unless $c^V_{ut} \approx c^A_{ut}$. This case, however, is not relevant for this work, since ALP-induced dark matter-nucleon scattering vanishes for $|c^V_{ut}| = |c^A_{ut}|$.

For ALPs with masses $m_a < \muCPT$, hadronic interactions with mesons can be described in chiral perturbation theory (chPT). Focusing on $\bc_D-$induced interactions, the chiral Lagrangian reads~\cite{Bauer:2021mvw}
\begin{align} \label{eq:L-chPT-FV}
    \mL_{\rm chPT} & \supset \frac{f_\pi^2}{8} \Tr[(\bD^\mu \bSigma) (\bD_\mu \bSigma)^\dagger] + \frac{f_\pi^2}{4} B_0 \Tr[\bmm_q (\bSigma^\dagger + \bSigma)],
\end{align}
with the pion decay constant $f_\pi \approx 130.5 \,\MeV$, the diagonal $3\times 3$ quark-mass matrix $\bmm_q$, the chPT parameters $B_0m_u = (6.2\pm 0.4)\cdot10^{-3} \,\GeV^2$, $B_0m_d \approx (13.3\pm 0.4)\cdot10^{-3}\,\GeV^2$~\cite{Bishara:2016hek}, and the covariant derivative of the chiral multiplet
\begin{align}
    \bD_\mu \bSigma = \partial_\mu \bSigma - i\, \frac{\partial_\mu a}{f_a}\bm k_D \bSigma\,,
\end{align}
where
\begin{align} \label{eq:chpt-Sigma}
    \bSigma = \exp[\frac{2 i }{f_\pi}\mqty( \frac{\pi^0}{\sqrt 2} + \frac{\eta_8}{\sqrt 6} & \pi^+ & K^+\\
    \pi^- & - \frac{\pi^0}{\sqrt 2} + \frac{\eta_8}{\sqrt 6} & K^0\\
    K^- & \bar K^0 & - \frac{2\eta_8}{\sqrt 6}
    )], \quad \bm k_D = \mqty(0 & 0 & 0 \\
    0 & 0 & (\bc_D)_{12}\\
    0 & (\bc_D)_{21} & 0
    ).
\end{align}
Here the couplings $(\bc_D)_{12}$ and $(\bc_D)_{21}$ are evaluated at the electroweak scale $\mu_W \approx m_t$. Assuming real ALP couplings and expan\-ding~\cref{eq:L-chPT-FV} to first order in the ALP couplings and ALP interactions with up to two meson fields, we obtain
\begin{align}\label{eq:chpt-kaons}
    \mL_{\rm chPT} & \supset - \frac{f_\pi}{\sqrt{2}f_a} (\bc_D)_{12} \,\partial^\mu a\,\partial_\mu K_L^0 \nn
    &\quad + \frac{i}{2} \frac{(\bc_D)_{12}}{f_a} \,a \left[K_S^0\, \partial^2 \left(\pi^0 - \sqrt{3} \,\eta_8 \right) - \left(\pi^0 - \sqrt{3} \,\eta_8 \right) \partial^2 K_S^0 \right] \nn
    &\quad + \frac{i}{2} \frac{(\bc_D)_{12}}{f_a} \,a \left[K^- \partial^2 \pi^+ - \pi^+ \partial^2 K^- \right] + h.c., 
\end{align}
with $K^0_L=(K^0 + \bar K^0)/\sqrt 2$ and $K^0_S=(K^0 - \bar K^0)/\sqrt 2$. The first term induces kinetic mixing between the ALP and $K_L^0$, through which the ALP can decay into three pions.\footnote{Mass mixing between the ALP and the pseudo-scalar mesons is absent as long as $c_{GG}=0$.} The terms in the second and third lines induce two-body decays of the ALP into light mesons. In particular, the decay width for $a\to K \pi$ is given by
\begin{align}\label{eq:alp-to-kpi}
    \Gamma(a\to K \pi) = \frac{m_a}{64\pi}\,|(\bc_D)_{12}|^2\frac{\qty(m_K^2 - m_{\pi}^2)^2}{f^2_a m_a^2}\,\lambda^{1/2}\qty(\frac{m_\pi^2}{m_a^2},\frac{m_K^2}{m_a^2})\,.
\end{align}
Summing up the three channels $K^0_S \pi^0,\,K^+ \pi^-,\,K^- \pi^+$ and neglecting mass differences, the decay length reads
\begin{align}
    \lambda_a &\approx \frac{0.6\,\rm mm}{|c^V_{ut} - c^A_{ut}|^2} \qty(\frac{f_a}{1\,\rm TeV})^2 \qty(\frac{m_a}{1\,\GeV})\lambda^{-1/2}\qty(\frac{m_\pi^2}{m_a^2},\frac{m_K^2}{m_a^2})\,.
\end{align}

To the best of our knowledge, this is the first time hadronic ALP decays through FC up-quark couplings are being investigated.

For ALPs with masses below the hadronic threshold, $m_\pi + m_e < m_a < 3 m_\pi$, semileptonic decay  $a\to \pi l \nu_l$, where $l=e,\mu$, is possible through kinetic mixing with $K^0_L$. For even lighter ALPs, decay into two photons is possible, but two-loop-suppressed. In both cases, the ALP is long-lived and can be considered as detector-stable.

\paragraph{ALP production and bounds}
Heavy ALPs with masses up to the TeV scale can be produced at the LHC in association with a top quark. For $m_a > m_t$, the dominant production channel is through the single-top process $pp \to tja$, where $j$ is a hadronic jet. For $m_a < m_t$, the ALP can also be produced from top decays in $pp\to t\bar t \to tja$. Both processes result in the same final-state particles, but with different kinematics. Depending on whether the ALP decays hadronically or invisibly, the signature consists of a reconstructed top quark and multiple jets or missing energy. In~\cite{Carmona:2022jid,Cheung:2024qve}, such signatures have been investigated for long-lived ALPs. The resulting bound of
\begin{align}\label{eq:cut-R-bound}
\frac{|(\bc_u)_{13}|}{f_a} = \frac{|c_{ut}^V + c_{ut}^A|}{2 f_a} \lesssim \frac{0.1}{\text{TeV}}
\end{align}
applies if the ALP has only couplings to right-handed up-type quarks and no flavor-diagonal couplings to quarks. 

ALPs with a coupling to left-handed quarks $(\bc_U)_{13} = (c_{ut}^V - c_{ut}^A)/2$ decay promptly through the induced down-quark couplings $(\bc_D)_{ij}$ from~\cref{eq:cD-ij}. In this case, measurements of top-quark kinematics can provide bounds on the ALP coupling, provided that the deviation from the Standard Model is moderate. In~\cite{Phan:2023dqw,Blasi:2023hvb}, kinematic distortions in top-antitop production have been investigated for ALPs with a flavor-diagonal coupling $c_{tt}$, yielding $|c_{tt}|/f_a \lesssim 10/$TeV. We expect a similar sensitivity to $(\bc_U)_{13}$ from distortions of the kinematic distributions in top-antitop production or from measurements of the chromomagnetic moment of the top quark. Other high-energy observables like the top-quark width or loop-induced ALP production from Higgs decays lead to even weaker bounds on $c_{ut}^V$ and $c_{ut}^A$~\cite{Carmona:2022jid,Bauer:2020jbp}. A dedicated search for $pp\to tja$ with hadronic ALP decays, for instance via $a\to b \bar d,\bar b d$ with bottom jets in the final state, could enhance the sensitivity to $(\bc_U)_{13}$.

Similar to flavor-diagonal couplings, ALPs with masses $m_a \lesssim 5\,$GeV can be produced in meson decays $B\to K a$, $B\to \pi a$ or $K\to \pi a$, if kinematically accessible. However, as $c_{ut}^{V} - c_{ut}^A$ induces flavor-changing down-quark couplings at tree level, ALPs with FC couplings can therefore be produced at high rates in meson decays. The branching ratios for $B^-\to K^- a$ and $B^- \to \pi^- a$ are~\cite{Ferber:2022rsf,Bauer:2021mvw}
\begin{align}
    \mathcal{B} (B^- \to K^- a) &\approx 2.6\cdot 10^3\, \abs{c_{ut}^V - c_{ut}^A}^2
    \qty(\frac{1\,\TeV}{f_a})^2  \frac{f_0^2(m_a^2)}{f_0^2(0)} \lambda^{1/2} \qty(\frac{m_K^2}{m_B^2},\frac{m_a^2}{m_B^2}),\nn
    \mathcal{B} (B^- \to \pi^- a) &\approx 6.5\cdot 10^4\, \abs{c_{ut}^V - c_{ut}^A}^2
    \qty(\frac{1\,\TeV}{f_a})^2  \frac{f_{0,B\to \pi}^2(m_a^2)}{f_{0,B\to \pi}^2(0)} \lambda^{1/2} \qty(\frac{m_\pi^2}{m_B^2},\frac{m_a^2}{m_B^2}),
\end{align}
where we have used the scalar hadronic form factors $f_0(q^2)$~\cite{Gubernari:2018wyi} and $f_{0,B\to \pi}(q^2)$~\cite{FermilabLattice:2015mwy}. To obtain these branching ratios, we have normalized the partial decay widths to the measured $B$ meson width, $\Gamma_{B}^{\rm exp}$~\cite{ParticleDataGroup:2024cfk}.\footnote{Numerically the branching ratios are much larger than those in~\cite{Ferber:2022rsf,Bauer:2021mvw}, as with FC ALP couplings the decay is induced at tree level, while FD couplings induce $B\to K$ and $B\to \pi$ through electroweak loops.} Notice that the branching ratio for $B\to \pi a$ is larger than for $B\to K a$, due to the CKM structure of the induced down-quark couplings, see~\cref{eq:cD-ij}. To probe $c_{ut}^{V,A}$, it can therefore be beneficial to explore $B\to \pi$ transitions in addition to the widely studied $B\to K$ transitions.
 
The most inclusive bound is obtained from the total width of the $B$ meson. Demanding that $\mathcal{B}(B^- \to \pi^- a) < 1$, we obtain the bound
\begin{align}\label{eq:cut-bound-bwidth}
   m_a < m_B - m_\pi: \quad \frac{|c_{ut}^V - c_{ut}^A|}{f_a} \lesssim \frac{0.004}{\TeV}\,.
\end{align}
For couplings $|c_{ut}^V - c_{ut}^A|/f_a \lesssim (0.024/\TeV) (m_a/\GeV)^{1/2}$, the proper decay length of the ALP exceeds $c\tau_a = 1\,$m and the ALP can be considered as long-lived at the Belle II  experiment. 
 In this case, Belle II's search for $B\to K\nu\bar\nu$ can be re-interpreted for an invisible intermediate ALP resonance. The latest re-interpretation has resulted in the bound $|(\bc_D)_{23}|/f_a \lesssim (0.2 \dots 2)\cdot 10^{-5}/$TeV~\cite{Abumusabh:2025zsr}, depending on the ALP mass, which translates to
\begin{align}\label{eq:cut-bound-btok}
   m_a < m_B - m_K: \quad \frac{|c_{ut}^V - c_{ut}^A|}{f_a} \lesssim \frac{(0.18 \dots 1.8)\cdot 10^{-4}}{\text{TeV}}\,.
\end{align}
Taken together, the measurement of the total $B$ meson width and the search for $B\to K \nu\bar\nu$ exclude the full range of coupling strength down to the bound of~\cref{eq:cut-bound-btok}.

For ALP masses $m_a < m_K - m_\pi$, fully hadronic decays are forbidden. Through kinetic mixing in chiral perturbation theory~\eqref{eq:chpt-kaons}, the ALP inherits the decay modes of the $K_L^0$, quadratically suppressed by the mixing parameter as $(f_\pi/f_a)^2$. Other possible decay modes are two-loop-suppressed. Due to the resulting long lifetime, the ALP can be considered as stable at fixed-target experiments and NA62's search for $K^+ \to \pi^+ X$ strikes again~\cite{NA62:2025upx}. Using the partial decay rate for $K\to \pi a$ from~\cite{Bauer:2021mvw} and the FC ALP coupling from~\cref{eq:cD-ij}, we derive the bound
\begin{align}\label{eq:cut-bound-ktopi}
   m_a < m_K - m_\pi: \quad \frac{|c_{ut}^V - c_{ut}^A|}{f_a} \lesssim \frac{(4.9 \dots 15.5) \cdot 10^{-8}}{\text{TeV}}\,,
\end{align}
which applies for ALP masses away from the pion mass.

\section{ALP-mediated dark matter-nucleus scattering}
\label{sec:dark matter-nucleus}
At direct detection experiments, the typical momentum exchange in dark matter-nucleus scattering, $|\bq| \lesssim 200\,$MeV, is small compared to the mass of the nucleus and the dark matter particles in the experimentally accessible range. Dark matter-nucleus interactions are therefore conveniently described within an EFT framework. We start with the Lagrangian for the ALP effective theory~\eqref{eq:lagrangian} at the cutoff scale $\Lambda$, evolve the theory down to the scale of chiral symmetry breaking, $\muCPT \sim 2$ GeV, where we match it onto the chiral effective theory that describes non-perturbative dark matter-nucleon scattering via ALP exchange. As the ALP must be integrated out of the theory around its mass scale, we present this procedure separately for ALPs with mass $m_a > \muCPT$ (Section~\ref{sec:dark matter-nucleon-heavy}) and ALPs with $m_a < \muCPT$ (Section~\ref{sec:dark matter-nucleon-light}). In the non-relativistic regime at $|\bq| \ll \muCPT$, the dark matter-nucleon interactions are finally matched onto dark matter-nucleus interactions using nuclear response functions. This last step 
 is applicable to both heavy and light ALPs.

\subsection{Heavy ALP mediator} \label{sec:dark matter-nucleon-heavy}
To describe dark matter-nucleon interactions through heavy ALPs, we largely follow the formalism and notation from~\cite{Bishara:2016hek,Bishara:2017pfq}. 
 If $m_a \gg \muCPT$, the ALP can be integrated out before matching onto the chiral Lagrangian. In this case, the ALP couplings from \eqref{eq:lagrangian} induce dark matter interactions with quarks and gluons, which are described by an effective Lagrangian~\cite{Brod:2017bsw,Bishara:2017pfq}\footnote{Here and in what follows, the dots represent terms that do not contribute to ALP-mediated dark matter-nucleon scattering at the lowest order.}
\begin{align} \label{eq:DM-quark}
    \mL_{\chi q}(\muCPT < \mu < m_a) = C_G O_G + C_{\widetilde G} O_{\widetilde G} + \sum_{q=u,d,s} \qty(C^q_S O^q_S + C^q_P O^q_P) + \dots
\end{align}
with the operators
\begin{align} \label{eq:operators}
    O_{G} &= \frac{\alpha_s}{12\pi} \qty(\bar \chi \chi) \, G^{b,\mu\nu} G^b_{\mu\nu},\\
    O_{\widetilde G} &= \frac{\alpha_s}{8\pi} \qty(\bar \chi i \gamma_5 \chi) \, G^{b,\mu\nu} \widetilde G^b_{\mu\nu},\nn
    O^q_S &= m_q \qty(\bar \chi \chi) \qty(\bar q q),\nn
    O^q_P &= m_q \qty(\bar \chi i \gamma_5 \chi) \qty(\bar q i \gamma_5 q).\notag
\end{align}
To one-loop order in QCD, the Wilson coefficients $C_G, C_{\tilde G}, C^q_S,$ and $C^q_P$ are scale-inde\-pen\-dent, except for a threshold correction to $C_G$ and $C_{\tilde G}$ each time a quark $Q$ is integrated out at the scale $\mu_Q \approx m_Q$~\cite{Hill:2014yxa}
\begin{align}\label{eq:threshold-corr}
    C_G^{n_f}(\mu_Q) &= C_G^{n_f+1}(\mu_Q) - C^{Q,n_f+1}_S(\mu_Q),\nn
    C_{\tilde G}^{n_f}(\mu_Q) &= C_{\tilde G}^{n_f+1}(\mu_Q) + C^{Q,n_f+1}_P(\mu_Q).
\end{align}
The superscript $n_f$ denotes the number of active quark flavors in the effective theory; it will be suppressed in what follows. At the scale of chiral symmetry breaking, the Wilson coefficients can be directly expressed in terms of high-scale Wilson coefficients, yielding
\begin{align} \label{eq:coupling-run-ma-muCPT}
    C^{u,d,s}_S (\muCPT) &= C^{u,d,s}_S(m_a),\nn
    C^{u,d,s}_P (\muCPT) &= C^{u,d,s}_P(m_a), \nn
    C_G(\muCPT) &= C_G(m_a) - \sum_Q C_S^Q(m_a), \nn
    C_{\tilde G}(\muCPT) &= C_{\tilde G}(m_a) + \sum_Q C_P^Q(m_a)\,.
\end{align}
The sum over $Q$ includes all quarks with $\muCPT<m_Q<m_a$.

At energies below $\muCPT \sim 2\,$GeV, quarks and gluons are confined inside color-neutral hadrons. The effective Lagrangian for (relativistic) dark matter interactions with nucleons $N = p,n$ reads
\begin{align} \label{eq:dark matter-nucleon}
    \mathcal{L}_{\chi N}(\mu < \muCPT) = \sum_{N = p,n} \qty(C_S^N O_S^N + C_P^N O_P^N) + \dots\,,
\end{align}
with the operators for scalar and pseudo-scalar interactions,
\begin{align}
 O_S^N & = (\bar \chi \chi)(\bar N N)\,,\\\nonumber
 O_P^N & = (\bar \chi i\gamma_5 \chi)(\bar N i \gamma_5 N)\,.
\end{align}
The DM-quark and DM-gluon Wilson coefficients from~\eqref{eq:coupling-run-ma-muCPT} match to the effective DM-nucleon interactions in chPT from~\eqref{eq:dark matter-nucleon} at tree level, which yields~\cite{Bishara:2016hek,Bishara:2017pfq}
\begin{align}\label{eq:matching}
    C^N_S (q^2) &= C_{ G} F^N_{ G} (q^2) + \sum_{q=u,d,s} C^q_S F^{q/N}_S(q^2)\,,\\
    C^N_P (q^2) &= C_{\widetilde G} F^N_{\widetilde G} (q^2) + \sum_{q=u,d,s} C^q_P F^{q/N}_P(q^2)\,.\notag
\end{align}
Here and throughout, $q^\mu = (E_R,\bq)$ is the four-momentum exchanged between the dark matter particle $\chi$ and the nucleon $N$, with recoil energy $E_R$ and three-momentum $\bq$. We define $\bq=\bm{k'}-\bm{k}$, where $\bm{k}$ and $\bm{k'}$ are the three-momenta of the incoming and outgoing nucleon. The nucleon form factors $F^{q/N}_S, F^{q/N}_P, F^N_{ G}, F^N_{\widetilde G} $ describe the quark and gluon distributions inside the nucleons. They are defined as~\cite{Bishara:2017pfq}
\begin{align} \label{eq:nucleon-FF}
    \bra{N'}{m_q \bar q q}\ket{N} &= F^{q/N}_S (q^2) \, \bar u_{N'} u_N,\\
   \bra{N'}{m_q \bar q i\gamma_5 q}\ket{N} &= F^{q/N}_P (q^2) \, \bar u_{N'} i \gamma_5 u_N,\nn
   \bra{N'}{ \frac{\alpha_s}{12\pi} G^{b\mu\nu} G^b_{\mu\nu} }\ket{N} &= F^{N}_{ G} (q^2)\, \bar u_{N'} u_N,\nn
   \bra{N'}{ \frac{\alpha_s}{8\pi} G^{b\mu\nu} \widetilde G^b_{\mu\nu} }\ket{N} &= F^{N}_{\widetilde G} (q^2)\, \bar u_{N'} i \gamma_5 u_N\,,\notag
\end{align}
where the nucleons are described by Dirac spinors $u_N$, $u_{N'}$ and $N,N'$ refer to the same nucleon, but with different momentum. It is important to note that the matching in~\eqref{eq:matching} only applies for DM-quark \emph{contact} interactions, \emph{i.e.}, for ALP masses $m_a \gg \mu_{\rm{chPT}}$.

\subsection{Light ALP mediator}\label{sec:dark matter-nucleon-light}
If the ALP mass lies below or around the scale of chiral symmetry breaking, $m_a \lesssim \muCPT$, the ALP must be included as a particle in the chiral effective theory. Unlike in the case of a heavy ALP, dark matter-nucleon scattering through light ALPs cannot be described by a contact interaction and the procedure to calculate dark matter-nucleon interactions from~\cref{sec:dark matter-nucleon-heavy} cannot be used. Instead, one must evolve and match the ALP effective theory from~\eqref{eq:lagrangian} to the chiral effective theory of ALP-nucleon interactions and compute the dark matter-nucleon scattering in the presence of light ALPs. To our knowledge, dark matter-nucleon interactions through a light ALP have not been described in the literature, though they are mentioned in~\cite{Fitzpatrick:2023xks}. 

The chiral effective theory for ALP-nucleon interactions has originally been constructed in~\cite{Georgi:1986df}. We follow the notation of~\cite{Bauer:2021mvw}. Since the ALP has the same quantum numbers as the neutral pion, ALP-pion mixing plays a crucial role in determining the amplitude for dark matter-nucleon scattering.

Below the scale of chiral symmetry breaking, $\mu < \muCPT$, ALP-pion mixing is induced by both ALP-quark and ALP-gluon couplings. The pion and ALP interaction eigenstates $\pi^0,\,a$ are related to the mass eigenstates $\hat{\pi}^0,\,\hat a$ via~\cite{Bauer:2020jbp}\footnote{This relation holds for small mixing, $m_\pi^2/|m_\pi^2 - m_a^2| \lesssim f_a/f_\pi$. ALPs with masses close to the pion mass would induce large effects in well-tested pion interactions. Moreover, this relation is scheme-dependent; for details see Ref.~\cite{Bauer:2020jbp}.}
\begin{align}
    \pi^0 &= \hat{\pi}^0 + \mO \qty(\frac{f_\pi^2}{f_a^2}) \nn
    a &= \hat{a} - \frac{1}{2\sqrt 2}\frac{f_\pi}{f_a} \frac{m^2_{\pi}}{m^2_\pi - m^2_a}\, \Delta c_{ud}  \,\hat{\pi}^0 + \mO \qty(\frac{f_\pi^2}{f_a^2}),
\end{align}
where $f_\pi  \approx 130.5\,$MeV is the pion decay constant and
\begin{align}
    \Delta c_{ud} = c_{uu}(\muCPT) - c_{dd}(\muCPT) + 2 c_{GG} \frac{m_d - m_u}{m_d + m_u}\,.
\end{align}

The ALP-gluon coupling $c_{GG}$ is scale-independent at least to two-loop order in the gauge couplings~\cite{Bauer:2020jbp}. The ALP-quark couplings $c_{uu}(\muCPT)$ and $c_{dd}(\muCPT)$ are related to the couplings at the cutoff scale $\Lambda$ of the ALP effective theory through the renormalization group. We apply the analytic evolution formulae from~\cite{Bauer:2020jbp}. For $\Lambda = 4\pi\,$TeV and neglecting sub-leading contributions from electroweak ALP couplings,
we obtain
\begin{align}
    c_{uu} (\muCPT) &\approx 0.98\, c_{uu}-0.12\,c_{tt}-\qty[\,3.6\,c_{GG} + 1.2 \,c_{bb} + 1.8\, \qty(c_{dd} + c_{ss} + c_{cc}) ]\cdot 10^{-2},\nn
    c_{dd}(\muCPT) &\approx 0.98\, c_{dd} + 0.11\,c_{tt} -\qty[\,3.7\,c_{GG} + 1.3\, c_{bb} + 1.9\, \qty(c_{uu} + c_{ss} + c_{cc}) ]\cdot 10^{-2},\label{eq:running-couplings}
\end{align}
where the couplings on the right-hand side are defined at the scale $\mu = \Lambda$.

The interactions of ALPs and pions with nucleons are described by the chiral Lagrangian~\cite{Gasser:1987rb,Bauer:2021mvw}
\begin{align}\label{eq:L-chbaryon}
    \mL_{\cpt} = \bar \mN \qty(i \slashed{\bD} - m_N\bm 1 + \frac{g_A}{2} \gamma^\mu \gamma_5 \bu_\mu + \frac{g_0}{2} \gamma^\mu \gamma_5 \bu_\mu^{(s)}) \mN, 
\end{align}
where $\mN = \qty(p,n)^T$ is the nucleon iso-doublet, $m_N= m_p \approx m_n$ is the nucleon mass, and $g_A \approx 1.25$ and $g_0 \approx 0.44$ are the external iso-vector and iso-scalar couplings~\cite{Liang:2018pis,Bauer:2021mvw}.
 The various operators are defined by
\begin{align}
    i\bD_\mu\, \mathcal{N} &= i \qty(\partial_\mu + \bGamma_\mu)\,\mathcal{N} \nn
    i \bGamma_\mu &= \frac12 \qty[\bxi \qty(i \partial_\mu + \br_\mu) \bxi^\dagger + \bxi^\dagger \qty(i \partial_\mu + \bl_\mu) \bxi] + v^{(s)}_\mu \bm 1\nn
    \bu_\mu &= \bxi \qty(i \partial_\mu + \br_\mu) \bxi^\dagger - \bxi^\dagger \qty(i \partial_\mu + \bl_\mu) \bxi \nn
    \bu^{(s)}_\mu &= \frac{\partial_\mu a}{2f_a} \qty(c_{uu} + c_{dd}   + 2c_{GG} ) \bm 1
    \equiv 2 a^{(s)}\frac{\partial_\mu a}{f_a} \bm 1\,.
\end{align}
The matrix $\bxi(x)$ is defined such that $\bxi^2(x) = \exp[\frac{i\sqrt{2}}{f_\pi}\bsigma^a \pi^a (x)]$, with pion fields $\pi^a$ and Pauli matrices $\bsigma^a$. Here and in the remainder of this section, the ALP couplings $c_{uu}$ and $c_{dd}$ are evaluated at the scale $\muCPT$, see~\eqref{eq:running-couplings}.

The iso-vector chiral couplings of ALPs to nucleons are
\begin{align}\label{eq:r-mu-l-mu}
    \br_\mu &= \frac{\partial_\mu a}{f_a} \qty[\frac{(\bmc_u - \bmc_d)_{11}}{2} + \frac{c_{GG}}{2}\frac{m_d-m_u}{m_d+m_u} + \frac{m_a^2}{m_\pi^2 - m_a^2} \frac{\Delta c_{ud}}{4}] \bsigma^3 \equiv \rho\,\frac{\partial_\mu a}{f_a} \bsigma^3\nn
    \bl_\mu &= \frac{\partial_\mu a}{f_a} \qty[\frac{(\bmc_U - \bmc_D)_{11}}{2}  - \frac{c_{GG}}{2}\frac{m_d-m_u}{m_d+m_u} - \frac{m_a^2}{m_\pi^2 - m_a^2} \frac{\Delta c_{ud}}{4}] \bsigma^3 \equiv \lambda\, \frac{\partial_\mu a}{f_a} \bsigma^3 \,,
\end{align}
and the iso-scalar vector coupling is
\begin{align}
    v^{(s)}_\mu = \frac{\partial_\mu a}{2f_a} \qty[\frac{\qty(\bmc_u + \bmc_d)_{11}}{2} + \frac{\qty(\bmc_U + \bmc_D)_{11}}{2}] \equiv v^{(s)}\frac{\partial_\mu a}{2f_a}\,.
\end{align}
At leading order in $|\bq|/{f_a}$ and $|\bq|/f_\pi$,
 the chiral Lagrangian \eqref{eq:L-chbaryon} includes interactions between neutral pions and nucleons 
\begin{align}\label{eq:pion-nucleon-interactions}
    \mL_{\cpt} & \supset \frac{g_A}{\sqrt 2} \frac{\partial_\mu \pi^0}{ f_\pi} \qty(\bar p \gamma^\mu \gamma_5 p - \bar n \gamma^\mu \gamma_5 n)\,,
\end{align}
as well as interactions between ALPs and nucleons
\begin{align}\label{eq:ALP-nucleon-interactions}
    \mL_{\cpt} & \supset \frac{\partial_\mu a}{2 f_a} \left( \bar p \gamma^\mu \qty{ v^{(s)} + (\rho + \lambda) + \qty[g_A (\rho-\lambda) + 2 g_0 a^{(s)}] \gamma_5} p \right.\nn
    &\hspace{1cm} \left.+\, \bar n \gamma^\mu \qty{ v^{(s)} - (\rho + \lambda) + \qty[-g_A (\rho - \lambda) + 2 g_0 a^{(s)}] \gamma_5}  n\right)\\\nonumber
    & = \frac{\partial_\mu a}{2 f_a} \Big( \bar p \gamma^\mu \Big\{\qty(\bmc_u + \bmc_U)_{11} + \frac12 g_{pa} \gamma_5\Big\} p 
    +\bar n \gamma^\mu \Big\{ \qty(\bmc_d + \bmc_D)_{11} + \frac12 g_{na} \gamma_5\Big\}  n\Big),
\end{align}
with the couplings
\begin{align}
    g_{pa} &= g_0 \qty(c_{uu} + c_{dd} + 2c_{GG}) + g_A \frac{m_\pi^2}{m_\pi^2-m_a^2} \Delta c_{ud}\nn
    g_{na} &= g_0 \qty(c_{uu} + c_{dd} + 2c_{GG}) - g_A \frac{m_\pi^2}{m_\pi^2-m_a^2} \Delta c_{ud}\,.
\end{align}
Notice that~\eqref{eq:pion-nucleon-interactions} and \eqref{eq:ALP-nucleon-interactions} are given in terms of interaction states $\pi^0, a$. At leading order in $f_\pi/f_a$, the interactions with nucleons are identical to those of the mass states $\hat{\pi}^0,\hat{a}$.

Finally, the leading interactions of ALP and pion mass eigenstates with dark matter are described by
\begin{align}\label{eq:ALP-dm-interactions}
    \mL_{\cpt-\chi} = \frac{c_\chi}{2f_a} \qty(\partial^\mu \hat{a} - \frac{1}{2\sqrt 2}\frac{f_\pi}{f_a} \frac{m^2_{\pi}}{m^2_\pi - m^2_a} \,\Delta c_{ud}  \,\partial^\mu \hat{\pi}^0) \bar \chi\, \gamma_\mu\gamma_5 \chi\,.
\end{align}
The ALP-DM coupling $c_{\chi}$ is not renormalized, as long as dark-sector forces are weak.

In summary, ALP-pion mixing introduces interactions between pions and dark matter, see~\eqref{eq:ALP-dm-interactions}, as well as additional interactions between ALPs and nucleons, see~\eqref{eq:r-mu-l-mu}, both proportional to $\Delta c_{ud}$. In \cref{sec:tree}, we will use the interactions from~\eqref{eq:pion-nucleon-interactions} -- \eqref{eq:ALP-dm-interactions} to calculate the amplitude for ALP-mediated dark matter-nucleon scattering at tree level.

\subsection{Non-relativistic scattering}\label{sec:nr-scattering}
The non-relativistic (NR) limit of dark matter-nucleon scattering can be obtained by expanding the dark matter and nucleon currents in $|\bm{q}|/m$. For scalar and pseudo-scalar currents, this yields\footnote{We use a non-relativistic normalization of currents; relativistic normalization would imply an extra factor $2m$.} 
\begin{align}\label{eq:nr-expansion}
\bar u^{s'} u^s & \to \zeta^{s'\dagger} \zeta^s\\\nonumber
\bar u^{s'}(p') i\gamma_5 u^s(p) & \to i(\bm{S})_{s' s}\cdot\frac{\bm{p} - \bm{p'}}{m}\,,
\end{align}
with the spin matrix $(\bm{S})_{s' s} = \zeta^{s'\dagger} \bm{\hat{S}}\, \zeta^s$, where the spin operator $\bm{\hat{S}} = \bm{\sigma}/2$ is directly proportional to the Pauli matrices $\bm{\sigma}$. The two-component spinors $\zeta^s,\zeta^{s'}$ fulfill the relation $\zeta^{s'\dagger} \zeta^s = \delta_{s's}$ with $s,s' = \{1,2\}$. 
The Lagrangian for non-relativistic dark matter-nucleon scattering is then given by~\cite{Bishara:2017pfq,Anand:2013yka}
\begin{align}\label{eq:nr-lagrangian}
    \mL_{\NR} = \sum_{N=n,p} c^N_1 \zeta_\chi^\dagger\zeta_N^\dagger\mO^N_1 \zeta_\chi \zeta_N + c^N_6 \zeta_\chi^\dagger\zeta_N^\dagger\mO^N_6 \zeta_\chi\zeta_N + \dots,
\end{align}
with the relevant NR operators for elastic scattering 
\begin{align}
\mathcal{O}^N_1 & = \mathbb{1}_\chi \mathbb{1}_N\\\nonumber
\mathcal{O}^N_6 & = \qty[\bm{\hat{S}}_\chi\cdot\frac{\bm{q}}{m_N}]\,\qty[\bm{\hat{S}}_N\cdot\frac{\bm{q}}{m_N}].
\end{align}
For heavy ALPs, the coefficients in the NR Lagrangian~\eqref{eq:nr-lagrangian} can be expressed in terms of the relativistic dark matter-nucleon interactions from~\eqref{eq:matching}, %
\begin{align} \label{eq:coefficient-NR}
    c^N_1 &= C^N_S\\\nonumber
    c^N_6 &= \frac{m_N}{m_\chi} C^N_P\,.
\end{align}
For light ALPs, the non-relativistic interactions must be calculated from the ALP-nucleon and ALP-dark matter interactions in~\eqref{eq:ALP-nucleon-interactions} and~\eqref{eq:ALP-dm-interactions}. We will perform this calculation in~\cref{sec:tree}.

\subsection{Dark matter-nucleus scattering}
The differential cross section for dark matter-nucleus scattering can be expressed as~\cite{Anand:2013yka}
\begin{align}\label{eq:dark matter-nucleus-xs}
    \dv{\sigma}{E_R} = \frac{m_A}{2\pi v^2} \frac{4\pi}{2J_A + 1} \sum_{\tau,\tau'} \sum_{O} R_O^{\tau \tau'} W_O^{\tau \tau'} (\absq)\,,
\end{align}
where $E_R$, $m_A$, and $J_A$ are the recoil energy, mass, and spin of the nucleus with mass number $A$, and $v$ is the dark matter velocity in the Earth's rest frame. The nuclear response functions
\begin{align}
    W^{\tau\tau'}_O (\abs{\bq}) = \sum_J \bra{J_A} O_{J,\tau}\ket{J_A}\bra{J_A} O_{J,\tau'}\ket{J_A},\quad \tau,\tau' = \{0,1\}\,,
\end{align}
are explicitly given in~\cite{Anand:2013yka,Fitzpatrick:2012ix}. Here the multipole operators $O = \{M, \Sigma^{'}, \Sigma^{''},\dots\}$ project onto the charge and the spin components of the nucleus. The indices $\tau,\tau' = \{0,1\}$ denote the (strong) isospin components of the interaction, where $0$ stands for isoscalar and $1$ for isovector. In the isospin-conserving limit where dark matter interactions with protons and neutrons are identical, the spin-independent response functions of a nucleus with mass number $A$ and atomic charge $Z$ reduce to
\begin{align}
    W^{00}_M & = \kappa_A A^2F^2(|\bq|)\,,\qquad W^{11}_M \to \kappa_A (A-2Z)^2F^2(|\bq|)\\\nonumber
    W^{01}_M & = W^{10}_M = -\kappa_A A (A-2Z)F^2(|\bq|)\,,
\end{align}
with $\kappa_A = (2J_A+1)/(4\pi)$ and $C(J_A) = \kappa_A(J_A + 1)/J_A$. The nuclear form factor $F(|\bq|)$ parametrizes the decoherence in dark matter-nucleus scattering and is often approximated by the Helm form factor \cite{Helm:1956zz}. In the long-wavelength limit $|\bq| \to 0$, coherent scattering off all nucleons occurs and $F(0) = 1$.

For a spin-dependent response, the response functions in the isospin and long-wavelength limit read 
\begin{align}
    W^{00}_{\Sigma'} & = 4C(J_A)(S_p + S_n)^2,\qquad W^{11}_{\Sigma'} = 4C(J_A)(S_p - S_n)^2\\\nonumber
    W^{01}_{\Sigma'} & = W^{10}_{\Sigma'} = 4C(J_A)(S_p^2 - S_n^2)\,,
\end{align}
where $S_{p}$ and $S_n$ are the spin averages of the protons and neutrons inside the nucleus. For $^{131}$Xe, one has $S_p \approx -0.009$ and $S_n \approx -0.272$ \cite{Klos:2013rwa}. In the long-wavelength limit, the spin-dependent response functions are related by $W^{\tau\tau'}_{\Sigma''} = W^{\tau\tau'}_{\Sigma'}/2$.

The nuclear response functions in~\eqref{eq:dark matter-nucleus-xs} are weighted by the model-dependent factors $R^{\tau\tau'}_O$, which encode the non-relativistic dark matter interaction with the nucleons. In particular, relativistic scalar and pseudo-scalar interactions induce spin-independent and spin-dependent non-relativistic scattering through~\cite{Bishara:2017pfq,Anand:2013yka}
\begin{align}
    R^{\tau\tau'}_M &= c_1^\tau c_1^{\tau'} + \dots\nn
    R^{\tau\tau'}_{\Sigma''} &= \frac1{16} \frac{\bq^4}{m_N^4} c^\tau_6 c^{\tau'}_6 + \dots, \label{eq:nuclear-coeff-squared}
\end{align}
where the NR coefficients are expressed as linear combinations of $c^p_i$ and $c^n_i$ from~\cref{eq:nr-lagrangian},
\begin{align}
    c^0_i = \frac12 \qty(c^p_i + c^n_i) 
    \,,\qquad
    c^1_i = \frac12 \qty(c^p_i - c^n_i)\,.
\end{align}
In the long-wavelength limit $1/|\bq| \gg R_A$, where $R_A = 1.2 A^{1/3}\,$fm estimates the size of the nucleus, the differential cross section for specific scattering processes reduces to a compact analytical formula. For spin-independent interactions $c_1^N$, the differential dark matter-nucleus scattering cross section then reads
\begin{align}\label{eq:dxsect-scalar}
    \dv{\sigma_{\mathrm{SI}}}{E_R} \to \frac{m_A}{2\pi v^2} \qty(Zc_1^p + \qty(A-Z) c^n_1)^2\,.
\end{align}
To compare with experimental results, we also define the average spin-independent scattering cross section per nucleon
\begin{align}\label{eq:SI-nucleon}
    \sigma_{\mathrm{SI},N} = \frac{\mu_{\chi N}^2}{\pi}\qty(\frac{Z}{A} c^p_1 + \frac{A-Z}{A} c^n_1)^2,
\end{align}
with the dark matter-nucleon reduced mass $\mu_{\chi N} = m_N m_\chi/(m_N + m_\chi)$.

For spin-dependent interactions $c_6^N$, the differential dark matter-nucleus scattering cross section in the long-wavelength limit is 
\begin{align} 
    \dv{\sigma_{\mathrm{SD}}}{E_R} \to \frac{m_A}{4\pi v^2} \frac{J_A + 1}{J_A} \qty(\frac{m_A\, E_R}{m^2_N})^2  \qty(c_6^p S_p + c_6^n S_n)^2. \label{eq:dxsect-pseudo-scalar}
\end{align}
Notice that spin-dependent scattering is suppressed at small recoil energy, $E_R \to 0$.

The expected scattering event rate per unit detector mass can finally be computed using the differential cross section via
\begin{align}\label{eq:event-rate}
    \dv{R}{E_R} = \frac{\rho_\chi}{m_A m_\chi} \int_{v_{\min}} \dv{\sigma}{E_R}\, v f_\oplus(\bv)\, d^3 \bv\,,
\end{align}
where $\rho_\chi$ is the local dark matter density, $f_\oplus(\bv)$ is the dark matter velocity distribution in the Earth's rest frame~\cite{Baxter:2021pqo}, and $v_{\min} = \sqrt{m_A E_R/2}/\mu_{\chi A}$ with the reduced mass $\mu_{\chi A} = m_A m_\chi/(m_A + m_\chi)$.

\section{Tree level: spin-dependent scattering}\label{sec:tree}
Based on the formalism from \cref{sec:dark matter-nucleus}, we calculate the dominant contributions to non-relativistic dark matter-nucleus scattering through the exchange of heavy and light ALPs. As anticipated, ALP exchange at tree level induces purely spin-dependent scattering. In \cref{sec:loop}, we will analyze spin-independent, loop-induced scattering.

\subsection{Heavy ALP mediator}\label{sec:heavy-alp-tree}
\begin{figure}[t]
    \centering
    \includegraphics[width=0.32\linewidth]{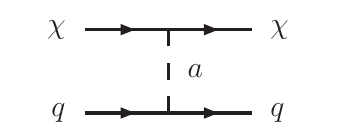}
    \includegraphics[width=0.32\linewidth]{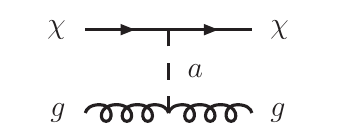}
    \includegraphics[width=0.32\linewidth]{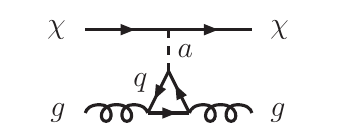}
    \caption{Feynman diagrams for 
     ALP-mediated dark matter-quark scattering (left) and dark matter-gluon scattering (center and right) at $\mathcal{O}(c^2/f_a^2)$.}
    \label{fig:LO-DM-heavy-ALP}
\end{figure}

At tree level, ALP exchange between the dark matter particles and the nucleon constituents, shown in fig.~\ref{fig:LO-DM-heavy-ALP}, generates pseudo-scalar interactions with quarks and interactions with the gluon configuration $G\widetilde{G}$. 
 We integrate out the ALP at the scale $\mu = m_a$. In terms of the effective ALP couplings from~\eqref{eq:lagrangian}, the Wilson coefficients from~\eqref{eq:DM-quark} read
\begin{align}\label{eq:heavy-alp-WC-at-ma}
    C_P^q (m_a) &= \frac{m_\chi}{m_a^2} \frac{c_\chi  c_{qq} (m_a)}{f_a^2}\,,\nn
    C_{\widetilde G}(m_a) &= -2\,\frac{m_\chi}{m_a^2} \frac{c_\chi }{f_a^2} \bigg[ c_{GG} + \frac12 \sum_{Q} c_{QQ} (m_a)\,\Theta(m_a - m_Q)\bigg]\,.
\end{align}
Here we have included the one-loop quark contribution to $C_{\widetilde G}$ (fig.~\ref{fig:LO-DM-heavy-ALP}, right), where the sum runs over all quarks $Q$ which are lighter than the ALP.

As mentioned in Sec.~\ref{sec:dark matter-nucleon-heavy}, the coefficients $C_P^q$ and $C_{\widetilde G}$ are not renormalized when evolved to lower scales~\cite{Hill:2014yxa}. The only effects are threshold corrections as the energy scale crosses the mass threshold of the various quarks, see \eqref{eq:coupling-run-ma-muCPT}. These threshold corrections remove the $c_{QQ}$ contribution of the respective quark from $C_{\widetilde G}$ in~\eqref{eq:heavy-alp-WC-at-ma}. Integrating out the charm quark at $\mu = \muCPT$, the low-scale Wilson coefficients for dark matter interactions  due to heavy-ALP exchange are given by
\begin{align}\label{eq:heavy-alp-tree}
    C_P^q (\muCPT) &= \frac{m_\chi}{m_a^2} \frac{c_\chi  c_{qq} (m_a)}{f_a^2}\,,\nn
    C_{\widetilde G}(\muCPT) &= -2\,\frac{m_\chi}{m_a^2} \frac{c_\chi }{f_a^2} \bigg[ c_{GG} + \frac12 \sum_{q=u,d,s} c_{qq} (m_a) \bigg]\,.
\end{align}
The scalar interaction $C_S^q$ is only generated at loop level, see~\cref{sec:loop}.
 Using the chiral matching conditions from~\eqref{eq:matching} and the relations from~\eqref{eq:coefficient-NR}, we match the effective couplings in~\eqref{eq:heavy-alp-tree} to the non-relativistic dark matter-nucleon interactions. The result is a spin-dependent NR interaction
\begin{align}\label{eq:sd-heavy-alp}
    c^N_6(q^2) = \frac{m_N}{m_a^2} \frac{c_\chi }{f_a^2} \bigg[\sum_{q=u,d,s} c_{qq}(m_a) F^{q/N}_P(q^2) - \bigg(2 c_{GG} + \sum_{q = u,d,s}\!\! c_{qq}(m_a)\bigg) F^N_{\widetilde G} (q^2)\bigg].
\end{align}
As the form factors $F^{q/N}_P(q^2)$ and $F^N_{\widetilde G} (q^2)$ are numerically of the same order, the dark matter-gluon interaction via a quark loop is of similar strength as the tree-level dark matter-quark interaction.

The cross section for ALP-mediated dark matter-nucleus scattering at tree level can finally be obtained by inserting $c^N_6$ from~\eqref{eq:sd-heavy-alp} into~\eqref{eq:dxsect-pseudo-scalar}.

\subsection{Light ALP mediator}\label{sec:light-alp-tree}
\begin{figure}[t]
    \centering
    \includegraphics[width=0.32\linewidth]{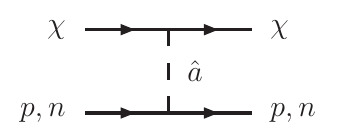}
    \includegraphics[width=0.32\linewidth]{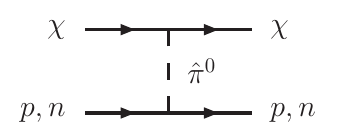}
    \caption{Feynman diagrams for tree-level ALP-mediated and pion-mediated dark matter-nucleon scattering at $\mathcal{O}(c^2/f_a^2)$.}
    \label{fig:tree-DM}
\end{figure}

For ALPs with masses $m_a < \muCPT$, dark matter can scatter with nucleons via the tree-level diagrams shown in \cref{fig:tree-DM}.
 The matrix element for relativistic dark matter scattering off protons via $\chi p \to \chi p$ is
\begin{align}\label{eq:nr-amplitude-tree}
    \mM_{\rm tree} &= - \frac{ g_{pa}}2 \frac{m_N m_\chi c_\chi}{f^2_a} \frac{1}{q^2 - m_{a}^2} \,\qty(\bar u_\chi i \gamma_5 u_\chi)\qty(\bar u_p i \gamma_5 u_p)\nn
    &\quad + \frac{g_A}{2}\frac{m_N m_\chi c_\chi }{f_a^2}\frac{1}{q^2 - m_{\pi}^2}\frac{m_\pi^2}{m_\pi^2 - m_a^2}\, \Delta c_{ud}\, \qty(\bar u_\chi i\gamma_5 u_\chi)\qty(\bar u_p i\gamma_5 u_p)\,.
\end{align}
The amplitude for scattering off neutrons, $\chi n \to \chi n$, is obtained by replacing $p \to n$ and $\Delta c_{ud} \to - \Delta c_{ud}$ in~\eqref{eq:nr-amplitude-tree}.
 This result is different from that of Ref.~\cite{Bae:2023ago}, where ALP-pion mixing was not considered. Matching~\eqref{eq:nr-amplitude-tree} onto the non-relativistic amplitude, we determine the spin-dependent dark matter-nucleon interaction in~\eqref{eq:nr-lagrangian} for protons and neutrons to be
\begin{align}\label{eq:spin-dep-light-alp}
    c^p_6(\bq^2) &= \frac{c_\chi}{2f_a^2}\,m_N^2 \qty[\frac{g_{pa}}{\bq^2 + m_{a}^2} - \frac{g_A}{\bq^2 + m_{\pi}^2}\frac{m_\pi^2}{m_\pi^2 - m_a^2} \, \Delta c_{ud} 
     ]\\
    c^n_6(\bq^2) &= \frac{c_\chi}{2f_a^2}\,m_N^2 \qty[\frac{g_{na}}{\bq^2 + m_{a}^2} + \frac{g_A}{\bq^2 + m_{\pi}^2}\frac{m_\pi^2}{m_\pi^2 - m_a^2} \, \Delta c_{ud} 
     ]\,. \notag
\end{align}
In the limit $m_a \to 0$, the interactions feature the typical $1/\bq^2$ behavior of a massless mediator.

\section{Loop level: spin-independent scattering}
\label{sec:loop}
Owing to the pseudo-scalar nature of the ALP, tree-level scattering is subject to both spin and momentum suppression. At the one-loop level, both types of suppression can be lifted through diagrams involving two ALP insertions. Such loop-induced processes generate \emph{scalar} dark matter-nucleon interactions, which are finite at small momentum exchange and induce coherent spin-independent scattering, but come at the price of a loop suppression. For heavy nuclei, coherent scalar interactions are enhanced by a factor of $A^2(m_N/\abs{\bq})^4$, see~\eqref{eq:nuclear-coeff-squared} and~\eqref{eq:dxsect-scalar}, compared to spin-dependent pseudo-scalar interactions. A priori, it seems plausible that loop-induced scattering could dominate over ALP-mediated tree-level scattering.

\begin{figure}[t!]
    \centering
    \includegraphics[width=0.32\linewidth]{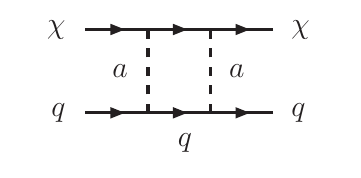}
    \includegraphics[width=0.32\linewidth]{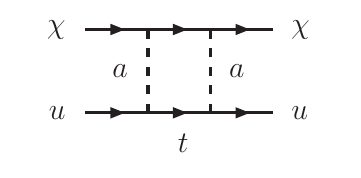}
    \caption{Feynman diagrams for loop-induced dark matter-quark scattering via flavor-diagonal (left) and flavor-changing (right) ALP couplings at $\mathcal{O}(c^4/f_a^4)$. Crossed diagrams also contribute to the scattering amplitude.
    \label{fig:loop-diagrams}}
\end{figure}

In this section, we will analyze three different classes of loop contributions that generate scalar dark matter-nucleon interactions. The first class, discussed in~\cref{sec:flavor-diagonal-loop}, involves flavor-diagonal ALP couplings to quarks. The corresponding loop diagram shown in~\cref{fig:loop-diagrams}, left, induces scalar interactions of dark matter with valence quarks and with the gluon condensate inside the nucleon.

The second class, discussed in~\cref{sec:flavor-changing-loop}, involves loops with flavor-changing neutral currents of ALPs coupling to virtual top quarks and valence up quarks, see~\cref{fig:loop-diagrams}, right. The large top mass leads to strongly enhanced scalar dark matter-nucleon interactions compared to flavor-diagonal contributions. For ALP masses $m_a > \muCPT$, both classes of one-loop amplitudes can be matched onto the effective Lagrangian~\eqref{eq:dark matter-nucleon} that describes dark matter-nucleon interactions at leading order in chiral perturbation theory.

In a third class of contributions, discussed in~\cref{sec:chiral-loop}, the spin suppression of tree-level scattering is lifted at loop level in chiral perturbation theory. Unlike in the previous cases, for ALP masses $m_a > \muCPT$ the \emph{tree-level} scattering amplitude is matched onto the effective Lagrangian~\eqref{eq:dark matter-nucleon}. The scalar dark matter-nucleon interaction is then generated by loop diagrams with virtual pion exchange, see~\cref{fig:loop-chiral-DM}.

As we will see, loop-induced dark matter-nucleon scattering through ALPs results in scattering rates that can be probed by current and upcoming direct detection experiments. Notice that scalar dark matter-nucleon scattering can also be generated by heavy mediators with masses above the cutoff scale of the ALP effective theory, see e.g.~\cite{Arcadi:2017wqi}. Such contributions are model-dependent, but could compete with the generic contributions discussed in this work. We will not include them in what follows. 

Throughout this section, we will assume that $m_a > \muCPT$, such that the ALP can be integrated out before matching onto the chiral Lagrangian. For $m_a < \muCPT$, the loop calculations would need to be carried out in the chiral effective theory of ALP-nucleon interactions described in~\cref{sec:dark matter-nucleon-light}. Phenomenologically, the case of light ALPs is less relevant, since strong experimental constraints on sub-GeV ALPs suppress dark matter-nucleon scattering way beyond observation at current and upcoming direct detection experiments.

\subsection{Flavor-diagonal ALP couplings} \label{sec:flavor-diagonal-loop}
For ALPs with flavor-diagonal couplings to quarks, we calculate the loop-induced amplitude for $\chi q \to \chi q$ scattering from the left diagram in~\cref{fig:loop-diagrams}, plus the corresponding diagram with crossed ALP couplings to the quark line. Only the axial-vector couplings of the ALP to up and down quarks, $c_{uu} = (\bmc_u - \bmc_U)_{11}$ and $c_{dd} = (\bmc_d - \bmc_D)_{11}$, contribute, see~\eqref{eq:ALP-quark-couplings}. The vector part $(\bmc_q + \bmc_Q)_{ii}$ does not contribute, because the remaining couplings involved in the amplitude conserve the flavor-diagonal vector current $\bar q \gamma_\mu q$. Our result for the relativistic one-loop scattering amplitude can be found in~\eqref{eq:full-amplitude-fd} of~\cref{app:analytics}.

For general external momenta, the amplitude contains only scalar and vector dark matter-quark interactions. Indeed, the two $\gamma_5$ insertions from the ALP-quark couplings prevent pseudo-scalar or axial-vector structures. The scalar contribution arises from the quark mass insertion $m_q$ on the propagator of the virtual quark in the loop, plus momentum contributions of external on-shell quarks, which reduce to $m_q$ when applying the Dirac equation. The same holds for the dark matter current. The vector contribution is generated only from momentum contributions that cannot be reduced by applying the Dirac equation on external states.

In the limit where the mass and momentum of the external quarks can be neglected with respect to the ALP mass, that is $m_q, |{\bf k}|, |{\bf k'}| \ll m_a$, we obtain the scattering amplitude in $D$ dimensions as
\begin{align}\label{eq:amplitude-FD}
    \mM_{\rm loop}^{\rm FD}(\mu) & = \frac{m_q m_\chi c^2_\chi\,c^2_{qq}(\mu)}{32 \pi^2 f_a^4}\, (\bar u_\chi u_\chi) (\bar u_q u_q) \bigg[\Delta - \ln(\frac{m_a^2}{\mu^2}) + L_{\rm FD}(m_a,m_\chi)\bigg],
\end{align}
with $q = \{u,d\}$ and the loop function
\begin{align}
    L_{\rm FD}(m_a,m_\chi) = 2 + \frac{(m_\chi^2 - m_a^2) \ln(m_a^2/m_\chi^2)}{m_\chi^2} - \frac{2m_a (3m_\chi^2 - m_a^2) \ln(\frac{m_a + \sqrt{m_a^2 - 4 m_\chi^2}}{2m_\chi})}{m_\chi^2 \sqrt{m_a^2 - 4 m_\chi^2}} \,.
\end{align}
The result is UV-divergent in 4 dimensions, as is apparent from $\Delta = 2/(4-D) - \gamma_E + \log(4\pi)$ and the remnant dependence on the renormalization scale $\mu$. We renormalize the amplitude in the MS--bar scheme at the scale $\mu = m_a$ by absorbing the divergence $\Delta$ into a $\bar\chi\chi\bar q q$ counterterm generated by the UV completion of the ALP effective theory~\eqref{eq:lagrangian}. We neglect potential model-dependent contributions that could be induced by heavy particles above the cutoff scale of the ALP effective theory.

For $m_a \gg m_q$, the loop-induced dark matter-quark scattering can be considered as a contact interaction and the formalism from~\cref{sec:dark matter-nucleon-heavy} applies. Matching the renormalized amplitude~\eqref{eq:amplitude-FD} onto the effective Lagrangian~\eqref{eq:DM-quark} at $\mu = m_a$ and evolving the Wilson coefficients down to the scale of chiral symmetry breaking using~\eqref{eq:coupling-run-ma-muCPT}, we obtain
\begin{align}
    C^q_S (\muCPT)&= \frac{m_\chi c^2_\chi c^2_{qq}(m_a)}{32 \pi^2 f_a^4} L_{\rm FD}(m_a,m_\chi)\qquad \text{for } q = \{u,d,s\} \,,\nn
    C_G(\muCPT) &= - \frac{m_\chi c^2_\chi}{32 \pi^2 f_a^4}\, L_{\rm FD}(m_a,m_\chi)\sum_{Q} c^2_{QQ}(m_a)\,.
\end{align}
The sum is over all quarks $Q$ with $\muCPT < m_Q < m_a$. Notice that $C_G$ is purely induced by quark threshold corrections, see~\eqref{eq:threshold-corr}. 
 We match these coefficients onto chiral perturbation theory using~\eqref{eq:matching} and obtain
\begin{align} \label{eq:nucleon-coupling-FD}
    C_S^N(q^2 < \mu_\cpt^2) = \bigg( \sum_{q = u,d,s}\!\!\! c^2_{qq}(m_a) F_S^{q/N}(q^2) -\sum_{Q} c^2_{QQ}(m_a) F_G^N(q^2) \bigg) \frac{m_\chi c^2_\chi}{32 \pi^2 f_a^4}\, L_{\rm FD}(m_a,m_\chi)\,,
\end{align}
where again $\muCPT < m_Q < m_a$.
In the non-relativistic limit $|\bq|/m_{\chi,N} \ll 1$, the Wilson coefficient can be readily identified with the coefficient for spin-independent scattering, $c_1^N = C_S^N$, see~\eqref{eq:coefficient-NR}.

In the limit of zero momentum transfer, $q^2 = 0$, the nucleon form factors are approximated by~\cite{Bishara:2017pfq}
\begin{align}
F_S^{q/N}(0) \approx (0.02 - 0.04)\, m_N\,,\quad F_G^N(0) \approx -0.06\,m_N\,,
\end{align}
where the quark form factor varies between 0.02 and 0.04, depending on the type of valence quark and nucleon. For similar ALP couplings $c_{qq} \approx c_{QQ}$, contributions from light and heavy quarks to the spin-independent scattering coefficient $c_1^N$ are numerically of the same order. Notice that both contributions add up constructively; cancellations between dark matter interactions with valence quarks and gluons do not occur.

In the limit $m_a \gg m_q,m_\chi$, loop-induced scalar interactions from FD ALP-quark couplings scale as $m_\chi m_N (m_{\chi}^2/m_a^2)/(32\pi^2 f_a^4)$. The decoupling for $m_\chi/m_a\to 0$ and the strong suppression by the cutoff scale of the ALP effective theory indicate a small contribution to the scattering rate.

Besides the dark matter-quark loop contributions from~\cref{fig:loop-diagrams}, effective ALP couplings to gluons can generate similar box diagrams to~\cref{fig:loop-diagrams}, left, where the quarks are replaced by gluons. In many models, effective ALP-gluon couplings are generated at the loop level, as suggested by the coupling definition in~\eqref{eq:lagrangian}. The resulting loop-induced dark matter-gluon interactions are therefore suppressed compared to the DM-quark interactions. We do not discuss loop-induced dark matter-gluon interactions in this work.

\subsection{Flavor-changing ALP couplings}\label{sec:flavor-changing-loop}
Compared to the case of flavor-diagonal couplings from~\cref{sec:flavor-diagonal-loop}, loop-induced dark matter-nucleon scattering with flavor-changing ALP couplings is strongly enhanced if the virtual quark in the loop is much heavier than the valence quarks inside the nucleon. In particular, for ALPs with flavor-changing couplings to up quarks and top quarks, the $\chi u \to \chi u$ scattering amplitude receives contributions with a virtual top quark from the right diagram in~\cref{fig:loop-diagrams}, plus the corresponding crossed diagram. Both the vector coupling $c_{ut}^V$ and the axial-vector coupling $c_{ut}^A$ of the ALP, defined in~\eqref{eq:FC-couplings}, contribute. The large top mass induces a quark chirality flip, which generates a scalar current scaling as $(m_t/f_a)\,\bar u u$. The amplitude for spin-independent dark matter-nucleon scattering through FC couplings is therefore enhanced by a factor $m_t/m_u$ compared to the FD case. This top-mass enhancement partially lifts the suppression by $f_a$. A similar mechanism has been observed for ALP-induced electromagnetic dipole moments with FC couplings to top quarks~\cite{DiLuzio:2020oah}.

For general external momenta, the dark matter-quark scattering amplitude includes all four types of Lorentz structures: scalar, pseudo-scalar, vector and axial-vector. However, only scalar and vector structures induce spin-independent coherent dark matter-nucleus scattering. We therefore neglect the pseudo-scalar and axial-vector interactions in this discussion. Our result for the full relativistic one-loop scattering amplitude can be found in~\eqref{eq:full-amplitude-fc} of~\cref{app:analytics}.

For external momenta much smaller than the top-quark mass as well as the ALP mass, $|{\bf k}|,|{\bf k'}| \ll m_t,m_a$, and neglecting contributions of $\mathcal{O}(m_u/m_t)$, the amplitude for $\chi u \to \chi u$ scattering through FC ALP couplings simplifies to
\begin{align}\label{eq:amplitude-FC}
    \mM_{\rm loop}^{\rm FC}(\mu)
    & = \frac{m_\chi m_t}{64 \pi^2 f_a^4} c_{\chi}^2 \left[ \abs{c_{ut}^V}^2(\mu) - \abs{c_{ut}^A}^2(\mu) \right] \qty(\bar u_\chi u_\chi)\qty(\bar u_u  u_u) \\\nonumber
    & \quad \cdot \bigg[-\Delta + \ln(\frac{m_t^2}{\mu^2}) + L_{\rm FC}(m_\chi,m_a,m_t)\bigg],
 \end{align}
where
\begin{align}
    L_{\rm FC}(m_\chi,m_a,m_t) &= \frac1{(m_t^2 - m_a^2)^2}\bigg[ m_a^2 \left( 2m_t^2  - m_a^2 \right) b_0(m_a,m_\chi) - m_t^4\, b_0(m_t,m_\chi)  \nn
    &\qquad\qquad\qquad\qquad\, + m_a^4 (m_t^2 - m_a^2) \frac{d}{dm_a^2} \,b_0(m_a,m_\chi) \bigg]  + \ln(\frac{m_\chi^2}{m_t^2}),\nn
    b_0(m_i,m_\chi) &= 2 - \frac{m_i^2}{2m_\chi^2} \ln(\frac{m_i^2}{m_\chi^2}) + \frac{m_i \sqrt{m_i^2 - 4 m_\chi^2}}{m_\chi^2} \ln(\frac{m_i + \sqrt{m_i^2-4m_\chi^2}}{2m_\chi}),
\end{align}
with $i = \{t,a\}$. Notice that the amplitude in~\eqref{eq:amplitude-FC} is proportional to the combination of couplings $|c_{ut}^V|^2 - |c_{ut}^A|^2$. In the case of purely chiral ALP-quark couplings, where either $(\bc_U)_{13}\to 0$ or $(\bc_u)_{13}\to 0$, the momentum-independent amplitude from~\eqref{eq:amplitude-FC} vanishes. In that case, the scattering amplitude is lead by terms suppressed as $|{\bf k}|/m_t$, $|{\bf k'}|/m_t$ or $m_u/m_t$ compared to~\eqref{eq:amplitude-FC}, see~\cref{app:analytics}. We expect these contributions to be of similar order as the FD amplitude from~\eqref{eq:amplitude-FD}.

As in the flavor-diagonal case, the FC amplitude from~\eqref{eq:amplitude-FC} is UV-divergent. Also here, we renormalize the amplitude in the MS-bar scheme using a $\bar \chi \chi \bar u u$ counterterm. For ALP masses $m_a < m_t$, we choose $\mu = m_a$ as renormalization scale, as this is the scale where the $\bar \chi \chi \bar u u$ contact interaction is generated upon integrating out the virtual particles. For $m_a > m_t$, we choose $\mu = m_t$.

Let us focus here on the case $m_a > m_t$. Evolving the flavor-changing ALP couplings from the cutoff scale $\Lambda = 4\pi\,$TeV down to the matching scale $\mu = m_t$~\cite{Bauer:2020jbp}, we find
\begin{align}
    c_{ut}^V (m_t) &\approx \phantom{+} 0.984 \, c_{ut}^V (\Lambda) - 0.005 \, c_{ut}^A (\Lambda) \,, \nn
    c_{ut}^A (m_t) &\approx -0.005 \, c_{ut}^V (\Lambda) + 0.984 \, c_{ut}^A (\Lambda)\,.
\end{align}
At the scale $\mu = m_t$, we match the renormalized FC amplitude onto the effective Lagrangian for DM-quark interactions from~\cref{eq:DM-quark}.\footnote{To be precise, one would first integrate out the ALP and subsequently the top quark. We neglect this extra step and integrate out both particles at $\mu = m_t$, since the scale separation between $\mu = m_a < 4\pi f_a$ and $\mu = m_t$ is moderate. This allows us to express the ALP-quark couplings $c_{ut}^V(m_t)$ and $c_{ut}^A(m_t)$ in terms of ALP couplings at the cutoff scale $\Lambda$.} Following the procedure described in \cref{sec:dark matter-nucleon-heavy}, we calculate the Wilson coefficient for scalar dark matter-nucleon interactions by matching the effective DM-quark interaction onto the dark matter-nucleon effective theory at $\mu=\muCPT$ using \cref{eq:matching}. We find
\begin{align}\label{eq:FC-coupling}
    C^N_S(q^2 < \muCPT^2) = F^{u/N}_S(q^2) \,\frac{m_t}{m_u}\frac{m_\chi}{64 \pi^2 f_a^4}\, c_{\chi}^2 \left[ \abs{c_{ut}^V}^2(m_t) - \abs{c_{ut}^A}^2(m_t) \right] L_{\rm FC}(m_\chi,m_a,m_t).
\end{align}
As anticipated, the scalar interaction from FC ALP couplings is enhanced by $m_t/m_u$ compared to the FD case~\eqref{eq:nucleon-coupling-FD}. The coefficient for non-relativistic spin-independent scattering can be easily obtained by identifying $c_1^N = C_S^N$, see~\eqref{eq:coefficient-NR}.

In the limit $m_\chi \gg m_a,m_t$, the effective scalar coupling $C^N_S$ scales as $m_\chi \ln(m_\chi^2/m_t^2)$. Dark matter scattering through FC ALP couplings is therefore significantly enhanced at large dark matter masses. On the other hand, in the limit $m_a \gg m_\chi,m_t$ the scalar coupling scales as $\ln(m_a^2/m_t^2)$, yielding a mild logarithmic dependence on the ALP mass.

\subsection{Pion loops in chiral perturbation theory}
\label{sec:chiral-loop}
In the previous two sections, scalar dark matter-nucleon interactions were generated by matching the ALP effective theory onto chiral perturbation theory at the one-loop level. But a scalar contact interaction can also be generated from loop diagrams \emph{within} chiral perturbation theory. While the tree-level matching of an ALP-quark interaction onto the chiral Lagrangian generates a pseudo-scalar dark matter-nucleon interaction, at loop level one can construct a scalar interaction via the exchange of a neutral pion or eta meson. However, the price to pay for turning a pseudo-scalar interaction into a scalar interaction is a chiral loop suppression. As we will show in this section, this suppression renders loop contributions in chiral perturbation theory subdominant compared to the contributions discussed in the previous sections.

\begin{figure}[t]
    \centering
    \includegraphics[width=0.32\linewidth]{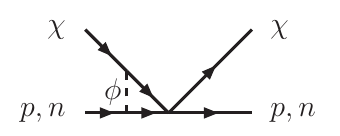}
    \includegraphics[width=0.32\linewidth]{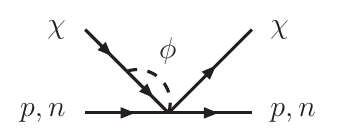}
    \includegraphics[width=0.32\linewidth]{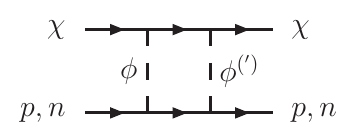}
    \includegraphics[width=0.32\linewidth]{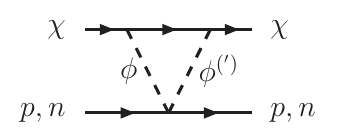}
    \includegraphics[width=0.32\linewidth]{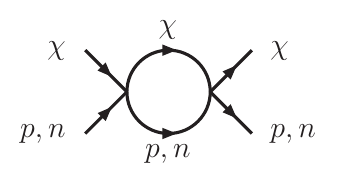}
    \caption{Feynman diagrams for loop-induced ALP-mediated dark matter-nucleon scattering with neutral meson exchange at $\mathcal{O}(c^4/f_a^4)$, where $\phi,\phi'=\pi^0,\eta$. Scalar $\chi \chi N N$ interactions are generated through various ALP-induced effective couplings in chiral perturbation theory: a $\chi\chi NN$ contact interaction (top left, bottom right), a $\chi\chi NN\phi$ contact interaction (top middle), two-meson exchange (top right), and a $NN\phi\phi^{(')}$  contact interaction (bottom left).    \label{fig:loop-chiral-DM}}
\end{figure}

In~\cref{fig:loop-chiral-DM}, we show Feynman diagrams in chiral perturbation theory that result in a scalar $\bar \chi \chi \bar N N$ interaction. The ALP is considered heavy compared to the scale of chiral symmetry breaking, $m_a > \muCPT$, leading to effective dark matter-hadron interactions and nucleon-meson interactions. To generate a scalar dark matter-nucleon interaction, the pion or eta meson must be emitted from the nucleon and connected to the dark matter line, in a way to ensure an even number of $\gamma_5$ insertions on the nucleon lines. In this section, we discuss only pion-mediated processes for simplicity. Diagrams involving the eta meson are enhanced by the large meson mass, but not enough to change the conclusions.

Since the energy scale of dark matter-nucleon scattering is much smaller than both the dark matter mass and the nucleon mass, one can work with an effective theory of heavy dark matter and heavy baryon interactions, formulated as the Heavy Dark Matter Effective Theory (HDMEFT) \cite{Hill:2011be,Hill:2013hoa,Hoferichter:2016nvd} and the Heavy Baryon Chiral Perturbation Theory (HBChPT) \cite{Georgi:1990um,Neubert:1993mb,Jenkins:1990jv,Falk:1990yz}, respectively.
 The main approach to HDMEFT or HBChPT is the same. The Dirac field $\psi$ of the heavy particle with mass $m$ and momentum $p = m v + \tilde p$, where $v^\mu$ is the four-velocity of the heavy particle and $\tilde p \ll m v$, is decomposed into particle and antiparticle components $h$ and $H$, so that
\begin{align}
    \psi(x) = e^{-im v x} \qty[h(x) + H(x)],
\end{align}
where
\begin{align}
    h(x) = e^{im vx} \frac{1 + \slashed v}{2} \psi(x)
    \,,\qquad
    H(x) = e^{im vx} \frac{1 - \slashed v}{2} \psi(x)\,.
\end{align}
At energies below the particle-antiparticle production threshold, one can integrate out the antiparticle component $H(x)$ to obtain an effective Lagrangian with the kinetic terms \cite{Neubert:1993mb,Mannel:1991mc}
\begin{align}
    \mL_{\rm eff} &= \bar h \,iv\cdot \partial\,h + \bar h\, i \slashed{\partial}_\perp \qty(i v \cdot \partial + 2 m - i \epsilon)^{-1} i\slashed{\partial}_\perp h \nn
    &= \bar h \,iv\cdot \partial\,h + \mO(\tilde p/m), \label{eq:HBChPT-Lagrange}
\end{align} 
where $\gamma^\mu_\perp = \gamma^\mu - v^\mu \,\slashed{v}$. Consequently, the propagator of the heavy Dirac field $h(x)$ is $i/v\cdot\tilde p$.
Without loss of generality, we choose the four-velocity of the heavy particle to be $v^\mu = (1,\bm{0})$~\cite{Neubert:1993mb}. We do this for both the dark matter particle and the nucleon, which is justified since the momentum exchange between them is of $\mathcal{O}(\tilde p/m)$. If the heavy particle is on-shell, its propagator is inversely proportional to its kinetic energy.

Integrating out the antiparticle modes $H$ in the interaction terms of the chiral Lagrangian, we have~\cite{Bishara:2016hek}\footnote{Compared to \cite{Bishara:2016hek}, we have added the interaction term $NN\pi\pi$ that can be derived from, for example, \cite{Bernard:1995dp}. Our convention for the transformation properties of the meson field matrix $\bSigma$, see \cref{eq:chpt-Sigma}, follows that of \cite{Bauer:2020jbp}, which differs from \cite{Bishara:2016hek} by hermitian conjugation. This results in a minus sign for the meson fields in the Lagrangian relative to \cite{Bishara:2016hek}.}
\begin{align}
    \mL &\supset \frac{C_{p\pi}}{f_\pi}\qty(\bar p\, i q\cdot S_p\, p ) \pi^0 - \frac{C_{p\pi\pi}}{2f_\pi} \qty(\bar p p)  \qty(\pi^0)^2 +  \qty(\frac{m_\pi^3}{m_\chi} C_{\chi \pi,0} - \frac{m_\pi}{m_\chi} q^2 C_{\chi \pi,1}) \qty(\bar \chi\,iq\cdot S_\chi\, \chi) \pi^0\nn
    &\quad + \frac{C_{\chi p}}{m_\chi} \qty(\bar\chi \, i q \cdot S_\chi \chi) \qty(\bar p \, i q \cdot S_p\,p) - \frac{C_{\chi p \pi}}{m_\chi} \qty(\bar\chi \, i q \cdot S_\chi\, \chi) \qty(\bar p p)\pi^0 + (p \to n,u\leftrightarrow d). \label{eq:L-chiral}
\end{align}
Here, we have defined $S^\mu = \gamma_\perp^\mu \gamma_5/2$. The coefficients are given by
\begin{align}\label{eq:chPT-coefficients}
    C_{p\pi} &= (-1)^{Q_p} \sqrt 2 (D+F),\nn
    C_{p\pi\pi} &= \frac{4}{f_\pi}\qty[b_0 m_d + \qty(b_0 + b_D + b_F) m_u],\nn
    C_{\chi \pi,0} &=\frac{B_0 f_\pi}{m_\pi^3\sqrt 2} \qty(m_u C^u_P - m_d C^d_P),\nn
    C_{\chi \pi,1} &= \frac{ \tilde m}{2 \sqrt 2} \qty(\frac1{m_u} - \frac1{m_d}) \frac{f_\pi}{m_\pi} C_{\tilde G},\nn
    C_{\chi p} &= \qty[D\qty(\frac{\tilde m}{m_u} + \frac{\tilde m}{m_s}) + F \qty(\frac{\tilde m}{m_u} - \frac{\tilde m}{m_s}) + G] C_{\tilde G},\nn
    C_{\chi p \pi} &= (-1)^{Q_p} \frac{2\sqrt 2 }{f_\pi} \qty[(b_0 + b_D + b_F) m_u C^u_P - b_0 m_d C^d_P],
\end{align}
with $\tilde m = \qty(1/m_u + 1/m_d + 1/m_s)^{-1}$ and the low-energy constants $D = 0.812(30),$ $F = 0.462(14)$, $G = -0.376(28)$, $B_0m_u = (6.2\pm 0.4)\cdot10^{-3} \,\GeV^2$, $B_0m_d \approx (13.3\pm 0.4)\cdot10^{-3} \,\GeV^2$, $b_0 = -3.7\pm 1.4$, $b_D = 1.4\pm 0.8$, $b_F = -1.8\pm 0.8$ \cite{Bishara:2016hek,Crivellin:2013ipa,GrillidiCortona:2015jxo}. The Wilson coefficients of the effective dark matter-gluon and dark matter-quark interactions, $C_{\tilde G}$, $C_P^u$ and $C_P^d$, were defined in~\cref{eq:heavy-alp-tree}. Using the central values for the low-energy constants, we have
\begin{align}\label{eq:chPT-coeff-num}
    C_{p\pi} &\approx -1.80 ,\nn
    C_{p\pi\pi} &\approx -0.81,\nn
    C_{\chi \pi,0} &\approx 0.23\ C^u_P(\muCPT) -0.50\ C^d_P(\muCPT) ,\nn
    C_{\chi \pi,1} &\approx 0.12\ C_{\widetilde G}(\muCPT),\nn
    C_{\chi p} &\approx 0.48\ C_{\widetilde G}(\muCPT),\nn
    C_{\chi p \pi} &\approx 0.20\ C^u_P(\muCPT) -  0.38\ C^d_P(\muCPT).
\end{align}
In this normalization convention, the Wilson coefficients of the chiral effective theory are either dimensionless constants of order one, or products of a dimensionless order-one constant and a Wilson coefficient of dark matter scattering with quarks and gluons. This choice allows us to separate the suppression of $1/(m_a^2 f_a^2)$
captured in the high-energy  coefficients $C^q_P$ and $C_{\widetilde G}$ from the momentum and pion-mass suppression in chiral perturbation theory. 

All diagrams shown in \cref{fig:loop-chiral-DM} are of order $\mathcal{O}(c^4/f_a^4)$, just as the perturbative one-loop contributions from~\cref{sec:flavor-diagonal-loop} and \cref{sec:flavor-changing-loop}. However, the chPT contributions are further suppressed by powers of $q/m_{\chi,N}$ and $m_\pi/m_{\chi,N}$.
To estimate this suppression, we introduce a 
 power counting scheme that takes account of the fact that all chPT coefficients in~\cref{eq:chPT-coefficients} are of similar magnitude. In the effective Lagrangian~\cref{eq:L-chiral}, we treat 
 the coefficients $C_{p\pi}, C_{p\pi\pi}$ as $\mO(1)$ and $C_{\chi\pi,0}, C_{\chi\pi,1},C_{\chi p}, C_{\chi p \pi}$ as $\mO(c^2/f_a^2)$,
 and count additional pre-factors as $m_\pi \sim f_\pi\sim q$.
  In this scheme, the coupling strength of the vertex $NN\pi$ is of order $\mO(1)$, $NN\pi\pi$ is of order $\mO(1/q)$, 
  $\chi\chi \pi$ is of order $\mO(q^4/m_\chi\times c^2/f_a^2)$, $\chi\chi NN$ is of order $\mO(q^2/m_\chi\times c^2/f_a^2)$, and $\chi\chi NN\pi$ is of order $\mO(q/m_\chi\times c^2/f_a^2).$
  This order-of-magnitude estimate 
  differs from the formal power counting often used in chiral perturbation theory, for instance in~\cite{Bishara:2016hek}.\footnote{In the formal counting scheme of chPT, only the particles' momenta and the quark masses $m_q \sim \mO(q^2/m_N)$ contribute to the power counting, while the potentially large low-energy constants are considered to be of $\mO(1)$. Consequently, the vertex $NN\pi$ is formally of $\mO(q)$, $NN\pi\pi$ and $\chi\chi NN$ are formally of $\mO(q^2),$ while $\chi\chi\pi$ and $\chi\chi NN\pi$ are formally of $\mO(q^3).$} Nevertheless, both power counting schemes give the same qualitative conclusion. 

In two-particle-irreducible diagrams, such as the second and fourth diagram in \cref{fig:loop-chiral-DM}, each loop integration yields a factor of $\mO(q^4/(16\pi^2))$, while each fermion or meson propagator contributes as $\mO(1/q)$ or $\mO(1/q^2)$, respectively. In two-particle-reducible diagrams such as the first, third and last diagram in \cref{fig:loop-chiral-DM}, the internal heavy fermion lines can be put on-shell, with a kinetic energy of order $\mO(q^2/m_{\chi,N})$. In that case, the power counting needs to be modified as suggested in \cite{Weinberg:1990rz,Epelbaum:2014efa,Korber:2017ery}: each nucleon or dark matter propagator that can be on-shell contributes as $\mO(m_{\chi,N}/q^2)$, each loop integral contributes $\mO(q^5/(16\pi^2m_{N}))$. 
 With these two kinds of counting, we determine the leading contribution of each diagram type in~\cref{fig:loop-chiral-DM} as
\begin{align}
    \text{Type 1 :} & \quad \frac{c^4}{f_a^4} \frac{q^5}{16\pi^2 m_\chi},\quad
    \text{Type 2 :}  \quad \frac{c^4}{f_a^4} \frac{q^6}{16\pi^2 m_\chi^2},\quad
    \text{Type 3 :}  \quad \frac{c^4}{f_a^4} \frac{q^5}{16\pi^2 m_\chi},\nn
    \text{Type 4 :} & \quad \frac{c^4}{f_a^4} \frac{q^6}{16\pi^2 m_\chi^2},\quad
    \text{Type 5 :} \quad \frac{c^4}{f_a^4} \frac{q^5}{16\pi^2 m_\chi}.
\end{align}
This estimate suggests that contributions of these chiral loop diagrams are negligible compared to the perturbative loop-level contributions in \cref{sec:flavor-diagonal-loop} and \cref{sec:flavor-changing-loop}. Therefore, we do not include them in our numerical analysis.

\subsection{Higher-order ALP couplings}\label{sec:ho}
All loop contributions considered in this work involve four effective ALP couplings at mass dimension 5, such that the amplitude scales as $c^4/f_a^4$. This raises the question if spin-independent scattering could also be generated by ALP couplings at higher orders in the ALP effective theory. In~\cite{Grojean:2023tsd}, an operator basis for all ALP couplings to SM particles up to mass dimension 8 has been constructed. With an intact shift symmetry, the only possible dimension-6 operator is a coupling of two ALPs to two Higgs fields. This coupling can indeed generate spin-independent scattering at loop level, with an amplitude scaling as $c^2/f_a^2$. For a generic pseudo-scalar, this contribution has been studied in~\cite{Ipek:2014gua}.

At mass dimension 7, all operators involve a single ALP field coupling to fermions through a partial derivative. This implies that no scalar interactions between fermion dark matter and quarks or gluons can be generated, because the ALP coupling to dark matter is momentum-suppressed. 

With the exception of the ALP-Higgs contribution, we do not expect any other contributions to spin-independent dark matter-nucleon scattering of order $1/f_a^4$ or larger in the ALP effective theory.

\section{Numerical results}
\label{sec:results}
Equipped with the analytic results on ALP-mediated dark matter-nucleon scattering at tree level (\cref{sec:tree}) and at one-loop level (\cref{sec:loop}), we perform a numerical analysis of the various contributions of ALP-mediated scattering to the event rates expected in direct detection experiments. Throughout our analysis, we focus on Xenon-based experiments. The current experiments XENONnT~\cite{XENON:2023cxc}, PandaX-4T~\cite{PandaX-4T:2021bab} and DarkSide-50~\cite{DarkSide:2022dhx} have world-leading sensitivity to dark matter-nucleon scattering for dark matter masses above a few GeV. The next-generation projects DARWIN/XLZD~\cite{Aalbers:2022dzr} and Pandax-xT~\cite{PANDA-X:2024dlo} have the ambition to probe dark matter-nucleon scattering down to cross sections comparable with atmospheric neutrino scattering.

To quantify ALP-mediated dark matter scattering at these experiments, we calculate the event rate $R$ from~\cref{eq:event-rate} and the related non-relativistic nucleus and nucleon scattering cross sections, $\sigma_{\rm SI}$ and $\sigma_{{\rm SI},N}$. The event rate $R$ depends on the local dark matter velocity distribution $f_\oplus(\bv)$, for which we implement the ``Standard Halo Model'' from Ref.~\cite{Baxter:2021pqo}.

The non-relativistic scattering cross sections involve QCD form factors for nucleons and nuclei. For the scalar nucleon form factors, we use the central values and uncertainties presented in \cite{Bishara:2017pfq}. For the pseudo-scalar and CP-odd gluonic form factors, we neglect the uncertainties, as their contributions to the event rates are strongly suppressed compared to the scalar contributions. For the nucleus form factors, we use the Mathematica script published in~\cite{Anand:2013yka}.\footnote{We rescale the spin-dependent nuclear response functions $W^{\tau\tau'}_{\Sigma''}$ obtained from the script to agree with the shell-model calculation~\cite{Klos:2013rwa} in the long-wavelength limit.
  We do not include uncertainties on the nuclear response functions. The spin-independent response functions computed by several independent groups \cite{Anand:2013yka,Hoferichter:2016nvd,Vietze:2014vsa} agree up to percent level in the relevant energy recoil range.} In our predictions, we take account of the relative abundances of various stable Xenon isotopes in the target material according to~\cite{Kondev:2021lzi}.

The hard scattering cross section is fully determined by the coupling and mass parameters of the ALP effective theory. Throughout our analysis, we set $f_a = 1\,$TeV, corresponding to a cutoff scale $\Lambda = 4\pi\,$TeV. We express all results for non-relativistic scattering in terms of ALP couplings defined at this cutoff scale, using renormalization group evolution as described before. The ALP-dark matter coupling always enters the observables in combination with ALP couplings to SM particles.
 In our numerics, we therefore quote the product of ALP-SM and ALP-DM couplings, for instance $c_{uu}c_\chi$. For the ALP couplings to SM particles, we consider one coupling at a time, setting all other couplings to zero. This allows us to analyze the individual effects of these couplings on dark matter-nucleon scattering. We comment on potential effects of additional couplings wherever they would alter the predicted cross sections.

As all loop-induced dark matter interactions are UV-sensitive, they depend on the choice of renormalization scale. In certain regions of parameter space, this scale dependence can lead to substantial variations of the scattering cross section. In our predictions, we include renormalization scale variations in the range $[0.5,2]\mu_R$, where $\mu_R$ is the central value chosen for the renormalization scale.

In what follows, we discuss the characteristics of the individual contributions to dark matter-nucleon scattering through ALPs with flavor-diagonal couplings (\cref{sec:FD-num}) and flavor-changing couplings (\cref{sec:FC-num}). In~\cref{sec:predictions}, we finally make predictions for the total scattering rate at Xenon-based experiments and discuss how they can probe the parameter space of the ALP effective theory.

\subsection{Flavor-diagonal ALP couplings}\label{sec:FD-num}
\begin{figure}[t]
    \centering
    \includegraphics[width=0.496\linewidth]{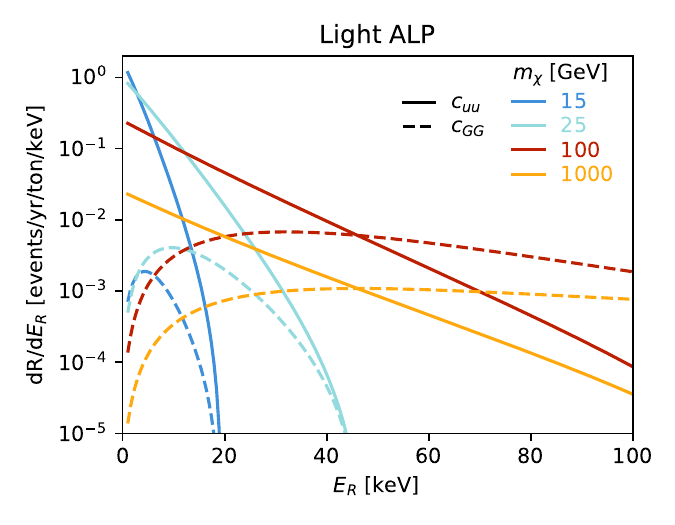}
    \includegraphics[width=0.496\linewidth]{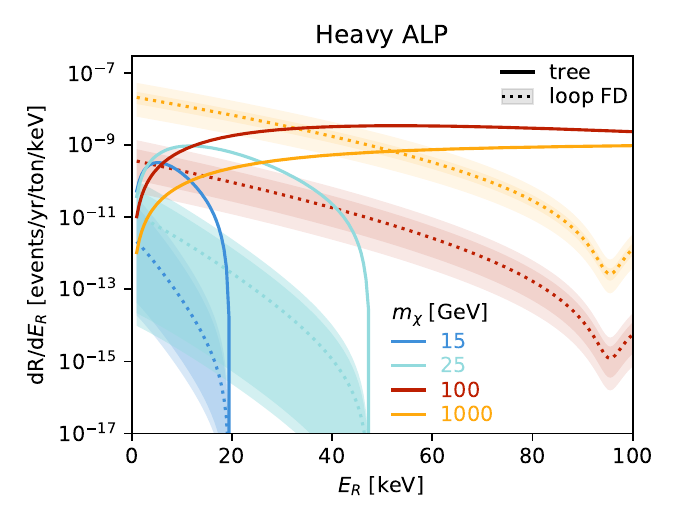}
    \caption{Energy recoil distribution $\dd R / \dd E_R$ of Xenon nuclei due to dark matter scattering through flavor-diagonal ALP couplings; shown for ALP masses $m_a = 1\,$MeV (left) and $m_a = 10\,$GeV (right) and
     various dark matter masses, with $f_a = 1\,$TeV.
      Left panel: spin-dependent tree-level scattering in two scenarios, $c_{uu}(\Lambda)c_\chi = 1,\,c_{qq \ne uu}(\Lambda) = c_{GG} (\Lambda) = 0$ (solid) and $c_{GG} (\Lambda)c_\chi = 1,\,c_{qq}(\Lambda) = 0$ (dashed). Right panel: spin-dependent tree-level (solid) and spin-independent one-loop (dotted) scattering for $c_{uu}(\Lambda)c_\chi = 1,\,c_{qq \ne uu}(\Lambda) = c_{GG} (\Lambda) = 0$. Colored bands indicate renormalization scale variation within $\mu \in [0.5 , 2]m_a$ (darker shade) and nucleon form factor uncertainties (lighter shade).
       \label{fig:ER-dist}}
\end{figure}

Flavor-diagonal ALP couplings to quarks generate spin-dependent dark matter-nucleus scattering at tree level and spin-independent scattering at one-loop level. In~\cref{fig:ER-dist}, we show the energy recoil spectra of Xenon nuclei induced by elastic dark matter scattering. We distinguish between light ALPs with masses $m_a < \muCPT$, represented by the benchmark $m_a = 1\,$MeV (left panel), and heavy ALPs with $m_a > \muCPT$, represented by $m_a = 10\,$GeV (right panel).

For light ALPs, we show spin-dependent dark matter-nucleus scattering through tree-level ALP and pion exchange, based on the amplitude derived in~\cref{sec:light-alp-tree}. The results are presented for two different scenarios with $c_{uu}(\Lambda)c_\chi = 1$ or $c_{GG} (\Lambda)c_\chi  = 1$ and all respective other couplings set to zero at the cutoff scale $\Lambda$. The difference between the spectra in both scenarios can be explained analytically in the long-wavelength limit. In this limit, the differential cross section takes on the form of~\cref{eq:dxsect-pseudo-scalar}, with the momentum-dependent non-relativistic couplings from~\cref{eq:spin-dep-light-alp}. The energy spectrum exhibits zeros when $c^p_6 S_p + c^n_6 S_n = 0$ or, equivalently, when
\begin{align}\label{eq:recoil-zeros}
    E_R = \qty[ \frac{g_A}{g_0} \frac{\Delta c_{ud}}{c_{uu} + c_{dd} + 2 c_{GG}} \frac{S_n - S_p }{S_n + S_p} - 1]  \frac{m_\pi^2}{2m_A}.
\end{align}
Here, all couplings are evaluated at $\muCPT$. For $c_{GG}(\Lambda)=1$, the solution~\eqref{eq:recoil-zeros} lies outside the physical region at $E_R \approx -2\cdot 10^{-4}$ keV,  manifesting itself as a decrease in the spectrum as $E_R \to 0$. The exact location of the cancellation point depends strongly on the momentum dependence of the form factor, as well as the values of the spin averages $S_n$ and $S_p$, all of which are affected by large uncertainties. Regardless, as long as $\abs{S_n} \gg \abs{S_p}$, which is generally true for Xenon isotopes with odd neutron numbers, the decrease close to $E_R = 0$ should persist. For $c_{uu}(\Lambda) = 1$, the spectrum exhibits a zero for $E_R \approx 139\,$keV. In this scenario, the energy spectrum is thus suppressed at large $E_R$, where it approaches the cancellation point. For light dark matter, the spectrum is cut off even earlier, as the maximum recoil energy is lower than for heavy dark matter.

For heavy ALPs, in~\cref{fig:ER-dist}, right, we display spin-dependent dark matter-nucleus scattering through tree-level ALP exchange (\cref{sec:heavy-alp-tree}) and spin-independent scattering through loop-induced FD ALP couplings to quarks (\cref{sec:flavor-diagonal-loop}), both in the scenario $c_{uu}(\Lambda)c_\chi = 1$. Unlike in the case of light ALPs, here the scattering proceeds through a contact interaction. The energy recoil spectrum is therefore determined by the nuclear form factors. Accidental cancellations in the recoil spectrum could only occur for scenarios with tuned values of $c_{qq}$ and $c_{GG}$, see~\cref{eq:sd-heavy-alp}.

As anticipated, the predicted event rates in both cases are small. With light ALP mediators, the momentum suppression of spin-dependent scattering at tree level is partially lifted and the rate is a priori within the current reach of XENONnT. However, as discussed in~\cref{sec:fd-bounds}, searches for $K^+\to \pi^+ X$ at NA62~\cite{NA62:2025upx} set strong bounds on $c_{GG}$ and $c_{qq}$. In~\cref{sec:predictions}, we will analyze the impact of these bounds on the predicted event rates.

With heavy ALPs, loop-induced spin-independent scattering through FD ALP couplings is similar in magnitude to spin-dependent scattering at tree level. At loop level, the momentum suppression is lifted, but this comes at the cost of a strong suppression by the cutoff scale of the ALP EFT, $f_a^{-4}$. This feature applies generically to ALP masses with $\muCPT < m_a \ll m_\chi$, for which the cross section has a mild, logarithmic dependence on ALP mass, see~\cref{eq:amplitude-FD}. With couplings of $\mathcal{O}(1)$, the resulting event rates are far below the reach of current experiments and lie inside the neutrino fog. We will not consider them any further in our analysis.

\subsection{Flavor-changing ALP couplings}\label{sec:FC-num}
\begin{figure}[t]
    \centering
    \includegraphics[width=0.496\linewidth]{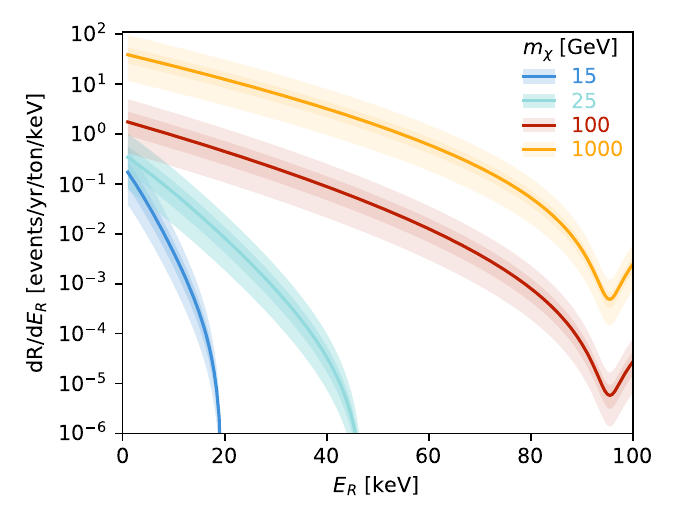}
    \includegraphics[width=0.496\linewidth]{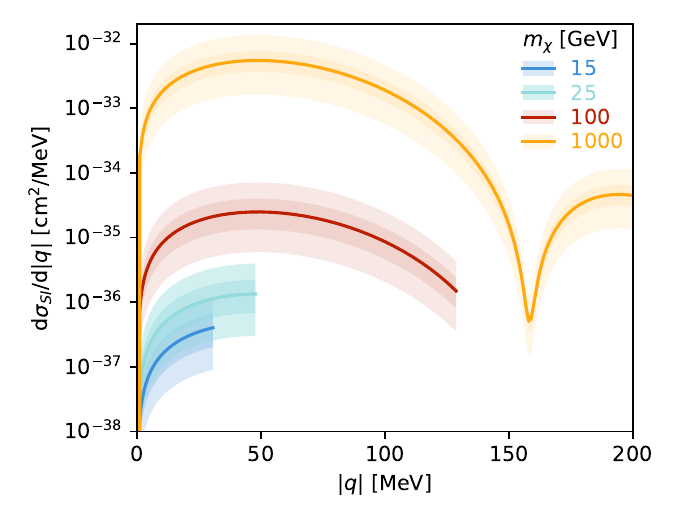}
    \caption{Energy recoil distribution $\dd R / \dd E_R$ (left) and momentum exchange distribution $\dd R / \dd \abs{q}$ (right) of Xenon nuclei due to dark-matter scattering through flavor-changing ALP couplings and various dark matter masses; shown for $m_a = 10\,$GeV and fixed ALP couplings $\bar c_{ut} c_\chi/f_a^2 = 1/$TeV$^2$. The momentum distribution is shown for fixed dark matter velocity $v = 340\,$km/s. Colored bands indicate renormalization scale variation within $\mu \in [0.5, 2]m_a$ (darker shade) and nucleon form factor uncertainties (lighter shade).
    \label{fig:FV-spectra}}
\end{figure}

Flavor-changing ALP couplings to quarks first contribute to dark matter-nucleus scattering at loop level, generating spin-independent interactions. The relevant scattering amplitude from~\cref{eq:amplitude-FC} probes the combination of ALP couplings $\bar c_{ut} c_\chi/f_a^2$, where we have defined
\begin{align}\label{eq:cut-bar}
\bar c_{ut} = \sqrt{\abs{|c_{ut}^V|^2(\Lambda) - |c_{ut}^A|^2(\Lambda)}}.
\end{align}
In~\cref{fig:FV-spectra}, we show the energy recoil distribution (left) and momentum exchange distribution for a fixed dark matter velocity of $v = 340\,$km/s (right) for the heavy-ALP benchmark with $m_a = 10\,$GeV. As for FD couplings, the shape of the spectra is determined by the nuclear form factors. In the right panel, the cutoff of the momentum distributions for light dark matter corresponds to the kinematic endpoint of momentum transfer in elastic scattering.

The recoil rate from spin-independent scattering through FC ALP couplings exceeds scattering through FD ALP couplings by 8 to 10 orders of magnitude, depending on the dark matter mass and recoil energy. As explained in~\cref{sec:flavor-changing-loop}, this large enhancement is due to the relative overall scaling of the scattering cross section with the quark mass ratio $(m_t/m_u)^2 \approx 10^{10}$. Moreover, for $m_\chi \gg m_a,m_t$, the scattering rate through FC ALP coup\-lings grows as $m_\chi^2\ln^2(m_\chi^2/m_t^2)$. The sensitivity of direct detection experiments to this scenario is thus particularly high for heavy dark matter.

As we will see in~\cref{sec:predictions}, the benchmark scenario from~\cref{fig:FV-spectra} and a large part of the ALP EFT parameter space can be probed at current experiments, provided that the dark matter mass is sufficiently large. We stress again that the top-mass enhancement of scattering through FC couplings requires ALP couplings to both left- and right-handed quarks to be present. Additional FD ALP couplings have little effect on these results, as the corresponding scattering amplitudes do not feature such an enhancement.

\subsection{Predictions for Xenon-based experiments}\label{sec:predictions}
\begin{figure}[t]
    \centering
    \includegraphics[width=0.496\linewidth]{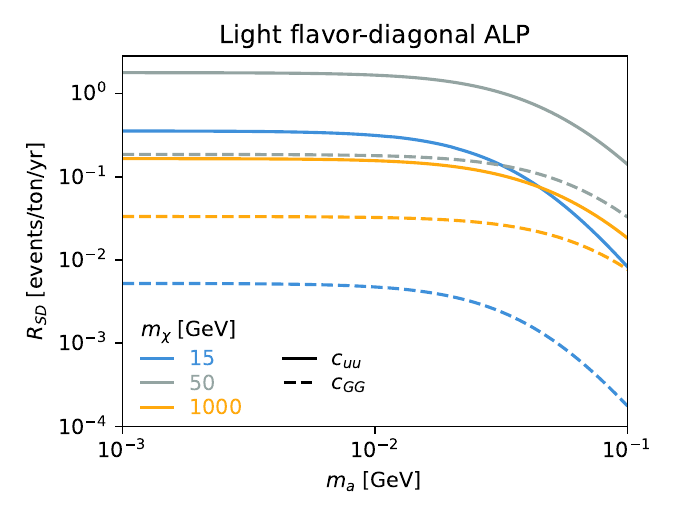}
    \includegraphics[width=0.496\linewidth]{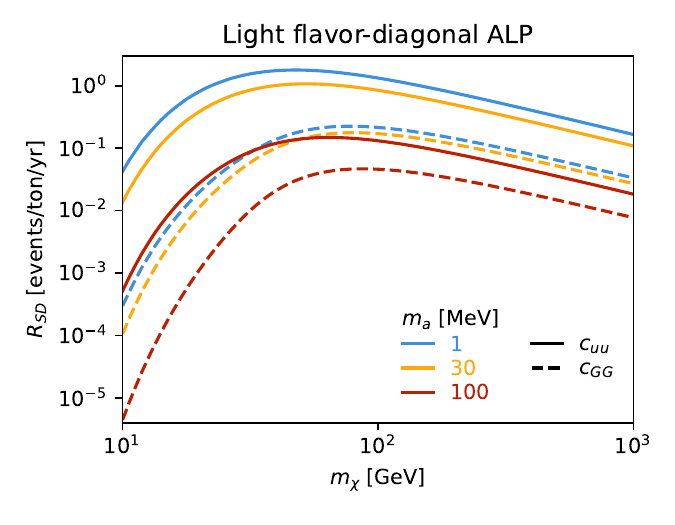}
    \includegraphics[width=0.496\linewidth]{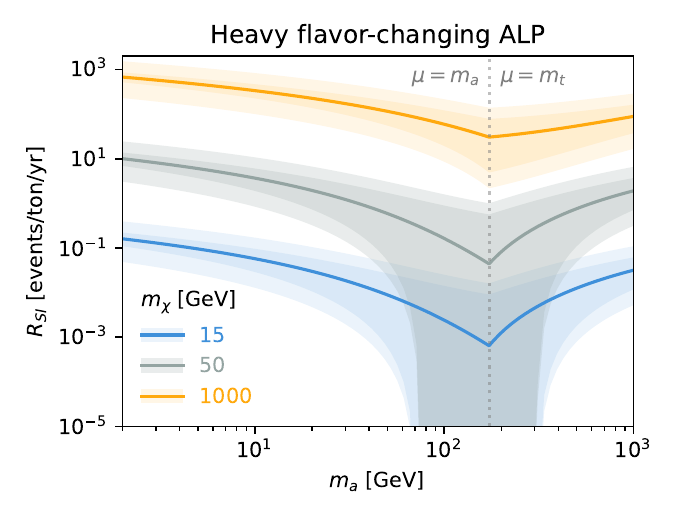}
    \includegraphics[width=0.496\linewidth]{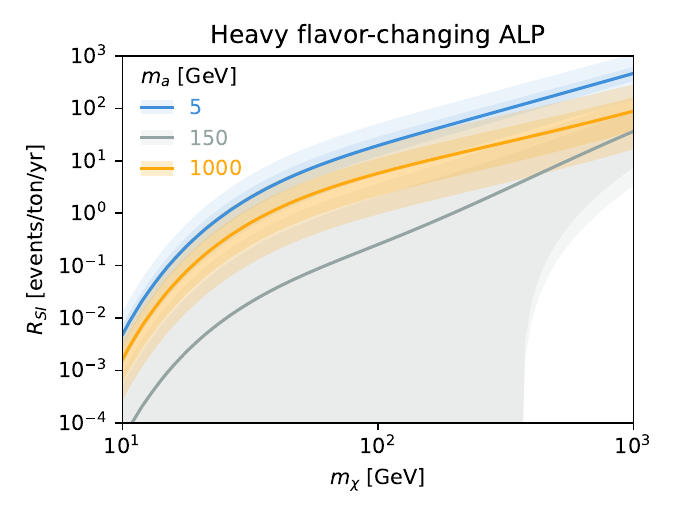}
    \caption{Total predicted event rate $R$ at the XENONnT experiment per ton-year. The detector efficiency and accessible energy recoil range described in~\cite{XENON:2025vwd} are taken into account. Upper row: spin-dependent scattering of light ALPs in two scenarios, $c_{uu}c_\chi/f_a^2 = 1/$TeV$^2$ (solid) and $c_{GG}c_\chi/f_a^2 = 1/$TeV$^2$ (dashed). Lower row: spin-independent scattering of heavy ALPs with flavor-changing couplings for $\bar c_{ut}c_\chi/f_a^2 = 1/$TeV$^2$. Colored bands indicate renormalization scale variation within $\mu \in [0.5, 2]m_a$ (darker shade) and nucleon form factor uncertainties (lighter shade).\label{fig:total-event-count}}
\end{figure}

By integrating the recoil spectra from the previous sections over the recoil energy, we obtain predictions for the expected event rates at Xenon-based direct detection experiments. In~\cref{fig:total-event-count}, we show these event rates for spin-dependent scattering through light ALPs with flavor-diagonal couplings (top row) and for spin-independent scattering through heavy ALPs with flavor-changing couplings (bottom row).

For ALPs with masses $m_a \lesssim 20\,$MeV and FD couplings, the predicted event rates for spin-dependent scattering approach the constant limit of an effectively massless ALP, see~\cref{fig:total-event-count} top left. In this region, the maximum event rates range around a few events per ton per year for ALP couplings $c_{uu}c_{\chi}/f_a^2 = 1/$TeV$^2$. However, the strong constraints from NA62's $K^+\to \pi^+X$ search on the ALP-quark coupling~\eqref{eq:cuu-bound-na62} and ALP-gluon coupling~\eqref{eq:cgg-bound-na62} suppress the rate by about 4-5 orders of magnitude. This is far beyond the reach of current and future direct detection experiments.\footnote{Notice that the bounds from $K^+\to \pi^+X$ are robust for ALP masses $m_a < 2 m_\mu$, even in the presence of ALP decays to electrons~\cite{NA62:2025upx}. At higher masses, resonance searches in $K^+\to \pi^+ \mu^+ \mu^-$~\cite{NA62:2025upx} and $B^+\to K^+\mu^+\mu^-$~\cite{LHCb:2015nkv} set strong constraints on $c_{uu}$ and $c_{GG}$, provided that the ALP branching ratio to leptons is substantial.}

For ALPs above the kaon mass, bounds on the couplings are much looser. However, for $m_a \gtrsim 20\,$MeV the scattering rate drops quickly because the $1/q^2$ enhancement of the ALP propagator at small momenta is cut off by the ALP mass. An observation of dark matter-nucleon  scattering through a light ALP at Xenon-based experiments is therefore extremely unlikely to occur.

On the contrary, the prospects to probe spin-independent scattering through ALPs with FC couplings are very promising. In the bottom panels of~\cref{fig:total-event-count}, we predict scattering rates of up to 1000 per ton-year for $m_{\chi} = 1\,$TeV. The growth of the scattering amplitude with the dark matter mass over-compensates the drop in the number density, such that the sensitivity is maximized for heavy dark matter. The ALP mass dependence is comparatively moderate. In~\cref{fig:total-event-count}, bottom left, the kink at $m_a = m_t$ is due to the change in renormalization scale. For $m_a < m_t$, one would first integrate out the top and choose the ALP mass as renormalization scale, and vice versa. The large scale dependence in this mass region is due to an accidental cancellation of the loop function for certain values of the mass parameters and renormalization scale. The strong renormalization scale dependence is an artifact of the UV sensitivity of the loop function, which originates from the missing heavy degrees of freedom in the ALP effective theory. 

\begin{figure}[t]
    \centering
    \includegraphics[width=0.6\linewidth]{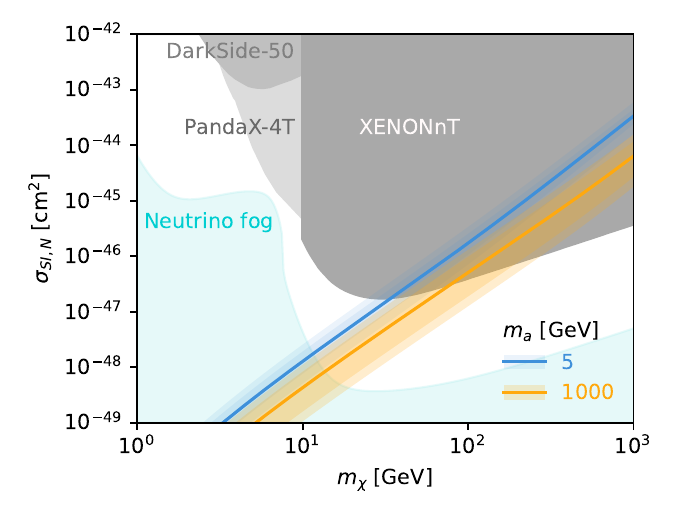}
    \caption{Cross section for spin-independent dark matter-nucleon scattering, $\sigma_{{\rm SI},N}$, versus dark matter mass, shown for different ALP masses and fixed couplings $\bar{c}_{ut} c_\chi/f_a^2 = 1/$TeV$^2$. Grey areas: experimental bounds from XENONnT~\cite{XENON:2025vwd}, PandaX-4T~\cite{PandaX:2023xgl} and DarkSide-50~\cite{DarkSide:2022dhx}. Blue area: neutrino fog \cite{OHare:2021utq}. Colored bands indicate renormalization scale variation within $\mu \in [0.5, 2]m_a$ (darker shade) and nucleon form factor uncertainties (lighter shade).
    \label{fig:xsect-vs-mDM}}
\end{figure}

To compare the predicted scattering rates against data, we calculate the average cross section for spin-independent dark matter-nucleon scattering from~\cref{eq:SI-nucleon}, $\sigma_{SI,N}$, which is reported by the experimental collaborations. In~\cref{fig:xsect-vs-mDM}, we show this cross section for two ALP mass benchmarks and fixed FC coupling $\bar c_{ut}c_{\chi}/f_a^2 = 1/$TeV$^2$. From the figure, it is apparent that XENONnT is already sensitive to these benchmarks for dark matter with masses $m_{\chi} \gtrsim 30\,$GeV. For light dark matter with $m_{\chi} \lesssim 10\,$GeV, the predicted cross section lies below the boundary of the neutrino fog and orders of magnitude below the sensitivity of current experiments. In this mass region, it will thus be very unlikely to observe ALP-mediated scattering at Xenon-based experiments.

\begin{figure}[t]
    \centering
    \includegraphics[width=0.6\linewidth]{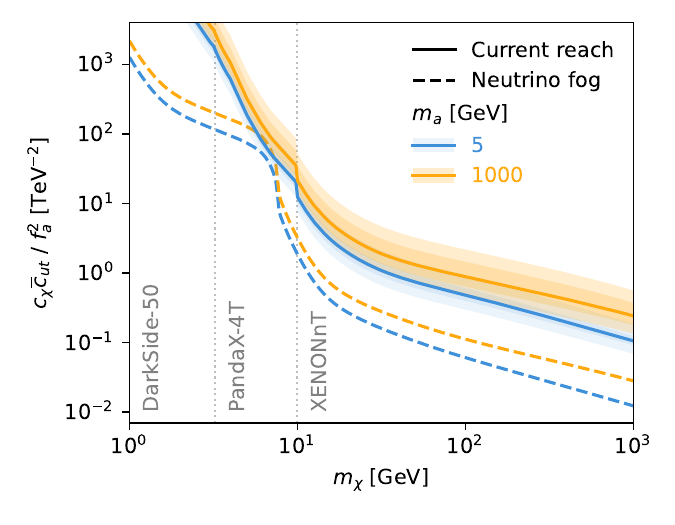}
    \caption{$90\%$ C.L. upper bounds on the combination of ALP couplings probed by dark matter-nucleon scattering, $\bar c_{ut} c_\chi/f_a^2$,
     versus dark matter mass, for different ALP masses.
 Solid lines: current upper bounds from XENONnT \cite{XENON:2025vwd}, PandaX-4T \cite{PandaX:2023xgl}, and DarkSide-50 \cite{DarkSide:2022dhx}. Dashed lines: projected bounds corresponding to cross sections at the neutrino fog boundary~\cite{OHare:2021utq}. Colored bands indicate renormalization scale variation within $\mu \in [0.5, 2]m_a$ (darker shade) and nucleon form factor uncertainties (lighter shade).
    \label{fig:cut-vs-mDM}}
\end{figure}

To quantify how much of the ALP parameter space can be probed with dark matter-nucleon scattering, in~\cref{fig:cut-vs-mDM} we show the current $90\%$ C.L. upper bounds on the product of ALP couplings, $\bar c_{ut}c_{\chi}$, at direct detection experiments. As anticipated, the sensitivity is particularly high at large dark matter masses. At small dark matter masses, direct detection experiments are only sensitive to flavor-changing ALPs if the couplings are close to the perturbative limit and/or if the cutoff scale of the ALP effective theory lies below the TeV scale.

A direct comparison of the reach at direct detection experiments with collider searches is limited for two reasons. First, dark matter scattering relies on the product $\bar c_{ut}c_{\chi}$, while colliders probe the ALP-quark coupling independently from the dark matter coupling. Second, dark matter scattering involves the combination $\bar c^2_{ut}\propto (\bc_u)_{13}(\bc_U)_{13}$ from~\cref{eq:cut-bar}, while the LHC bound $|(\bc_u)_{13}|/f_a\lesssim 0.1/$TeV from~\cref{eq:cut-R-bound} applies only if $(\bc_u)_{13} \gg (\bc_U)_{13}$. We therefore do not attempt a comparison with collider searches, which would only apply for a restricted part of the ALP parameter space. 

For heavy ALPs with sizeable couplings to left-handed quarks, direct detection experiments currently offer the best sensitivity to the FC coupling $\bar c_{ut}$ -- provided that the ALP also couples to the dark sector. As discussed in~\cref{sec:fc-bounds}, dedicated searches for top-associated ALP production with prompt hadronic decays, for instance through $pp\to tja,\,a\to j_bj$ with a bottom-quark jet $j_b$, would be required to probe this scenario at the LHC. Alternatively, FC ALP couplings could be probed in same-sign top production through $t-$channel ALP exchange. Such processes could be identified, for instance, through signatures with same-sign leptons~\cite{Arganda:2023pye}.

To estimate the reach of future direct detection experiments, we also show the ALP couplings corresponding to a cross section at the boundary of the neutrino fog. The anticipated sensitivity to the effective couplings increases by an order of magnitude. In the framework of the ALP effective theory, this means that dark matter-nucleon scattering will probe new physics beyond the TeV scale, provided that it generates sizeable flavor-changing couplings to ALPs. This is a remarkable result, given that the underlying scattering amplitude is generated at the loop level and the scattering occurs at very low energy scales, with a four-fold suppression by the cutoff scale $\Lambda \propto f_a$.

Let us close with a remark on light ALPs with masses $m_a < \muCPT$ and FC couplings. While such a scenario requires a dedicated calculation in chiral perturbation theory, we expect a priori sizeable scattering rates also in this case. This is suggested by the logarithmic growth of the event rate for small $m_a$, see the bottom left panel of~\cref{fig:total-event-count}. At colliders, this mass region is probed by the total $B$ meson width, as well as searches for invisible ALPs in $B\to K\nu\bar\nu$ and $K\to \pi X$, as discussed in~\cref{sec:fc-bounds}. Taken together, these observables constrain FC ALP couplings to $|c_{ut}^V - c_{ut}^A|/f_a < (0.004 \dots 5\cdot 10^{-8})/$TeV, depending on the ALP mass. These strong bounds make an observation at direct detection experiments very challenging, even for heavy dark matter where the scattering rate is maximized.

\section{Conclusion}\label{sec:conclusion}
With this systematic analysis of ALP-mediated dark matter-nucleon scattering, we have identified several qualitatively different contributions to spin-dependent and spin-indepen\-dent scattering. Spin-dependent scattering is generated at tree level through ALP exchange, but momentum-suppressed. A priori, this suppression can be lifted if the ALP is lighter than the typical momentum transfer. However, strong bounds on the ALP couplings to quarks and gluons from searches for invisible ALPs produced in kaon decays at NA62 suppress the predicted scattering rates far below the reach of current and future direct detection experiments.

Spin-independent scattering is generated through two kinds of loop processes. At energies above the scale of chiral symmetry breaking in QCD, double ALP exchange generates scalar interactions between dark matter and the quarks and gluons inside the nucleon. With flavor-diagonal ALP couplings, the scattering amplitude is suppressed as $c^4/f_a^4$, which leads to negligible event rates for experimental purposes. At low energies, loop exchange of virtual mesons and nucleons generates scalar interactions in chiral perturbation theory. However, these contributions are suppressed by several powers of the small momentum transfer, in addition to the $c^4/f_a^4$ suppression.

The phenomenology of ALP-mediated dark matter scattering completely changes if the ALP has flavor-changing couplings to top and up quarks. The large top mass enhances the loop-induced scattering amplitude by 5 orders of magnitude compared to contributions from ALPs with flavor-diagonal couplings, compensating for the loop and cutoff-scale suppression. Spin-independent dark matter-nucleon scattering through flavor-changing ALPs occurs at high rates, which are already probed by current Xenon-based experiments. The scattering cross section grows with the dark matter mass, so that event rates are highest for heavy dark matter, despite the comparatively lower number density. Model-dependent contributions from particles living above the cutoff scale of the ALP effective theory could further enhance the scattering rate.

The future direct detection experiments DARWIN/XLZD and PandaX-xT are expected to probe large parts of the ALP parameter space through spin-independent scattering. Our predictions suggest that their sensitivity to flavor-changing ALP couplings will exceed that of collider searches, provided that the ALP has sizeable couplings to dark matter. We encourage the experimental collaborations to consider ALP-mediated dark matter as an interesting target for their searches.

\section{Acknowledgment} We thank Yotam Soreq for bringing recent work on hadronic ALP decays to our attention.

\appendix
\section{Loop functions for spin-independent scattering}\label{app:analytics}
In this appendix, we provide the relativistic one-loop amplitudes for ALP-mediated dark matter-nucleon scattering. These results are valid for on-shell external states with arbitrary momentum and apply for any ALP mass above the scale of chiral symmetry breaking, $m_a > \mu_\cpt$.

\paragraph{Flavor-diagonal ALP couplings} For FD ALP-quark couplings, the one-loop amplitude for the scattering process $\chi(p) + q(k) \to \chi (p') + q(k')$ with momentum exchange $q = k' - k$ is
\begin{align} \label{eq:full-amplitude-fd}
    \mM_{\rm loop}^{\rm FD}  &= \frac{m_q m_\chi c^2_{qq}c^2_\chi}{16 \pi^2 f_a^4}  \bigg\{- m_\chi \qty(\bar u_\chi \gamma^\mu u_\chi) \qty(\bar u_q u_q) C_\mu (q,p,m_a,m_a,m_\chi)\nn
    &\quad + m_q \qty(\bar u_\chi u_\chi) \qty(\bar u_q  \gamma^\mu u_q) C_\mu (-k,q,m_a,m_q,m_a) + \frac12 \qty(\bar u_\chi u_\chi) \qty(\bar u_q  u_q) B_0 (q,m_a,m_a)\nn
    &\quad - m_\chi m_q \qty(\bar u_\chi \gamma^\mu u_\chi) \qty(\bar u_q  \gamma^\nu u_q) \Big[D_{\mu\nu} (-k,q,p,m_a,m_q,m_a,m_\chi)\nn
    &\qquad\qquad\qquad\qquad\qquad\qquad\quad - D_{\mu\nu} (k',q,p,m_a,m_q,m_a,m_\chi)\Big] \bigg\},
\end{align}
where we have used the shorthand notations
\begin{align}
    \bar u_\chi \equiv\bar{u}_\chi(p'),
    \quad
    u_\chi \equiv u_\chi(p),
    \quad
    \bar u_ q \equiv \bar u_q(k'),
    \quad
    u_q \equiv u_q(k).
\end{align}
The corresponding Feynman diagram is given in~\cref{fig:loop-diagrams}, left.
The scalar and tensor integral functions $B_0$, $C_\mu$ and $D_{\mu\nu}$ are defined as in~\cite{Denner:1991kt}. Effective scalar interactions $\qty(\bar u_\chi u_\chi) \qty(\bar u_q u_q)$ are generated by the $B_0$ and $C_\mu$ terms.
Effective vector interactions $\qty(\bar u_\chi \gamma^\mu u_\chi) \qty(\bar u_q  \gamma_\mu u_q)$ are due to the $D_{\mu\nu}$ terms. They are suppressed by at least a factor of $\mO(m_q/m_a)$ compared to the scalar interactions.

Our result in~\eqref{eq:full-amplitude-fd} differs from that obtained in~\cite{Ipek:2014gua,Arcadi:2017wqi}. The latter articles consider a generic pseudo-scalar mediator $P$ with a coupling $\bar\psi i\gamma_5\psi P$ to fermions $\psi$. In our case the pseudo-scalar ALP has derivative couplings to fermions, see~\cref{eq:lagrangian}, which makes a difference in the loop function. However, we are puzzled by the fact that the loop functions in~\cite{Ipek:2014gua} and~\cite{Arcadi:2017wqi} seem to disagree.

\paragraph{Flavor-changing ALP couplings} For FC ALP-top-up couplings, the one-loop scattering amplitude for $\chi(p) + u(k) \to \chi (p') + u(k')$ is
\begin{align} \label{eq:full-amplitude-fc}
    \mM^{\rm FC}_{\rm loop} &= -\frac{m_t m_\chi c_\chi^2}{64\pi^2 f_a^4}\bigg\{   \left( \bar{u}_\chi u_\chi \right) \left( \bar{u}_u \mC_{ut}^- u_u \right) \left[ B_0(q,m_a,m_a)  + m_t^2 C_0(k,-q,m_a,m_t,m_a) \right]\nn
    &+ m_\chi  \left( \bar{u}_\chi \gamma^\mu u_\chi \right) \left( \bar{u}_u   \mC_{ut}^-  u_u \right) \left[ -2 C_\mu (q,p,m_a,m_a,m_\chi) \right. \nn
    &\left.\quad + m_t^2 D_\mu (k,-q,-p,m_a,m_t,m_a,m_\chi) +  m_t^2 D_\mu (-k',-q,-p,m_a,m_t,m_a,m_\chi) \right] \nn
    &+ m_\chi m_t  \left( \bar{u}_\chi \gamma^\mu u_\chi \right) \left( \bar{u}_u    \gamma^\nu \mC_{ut}^+ u_u \right)[ D_{\mu\nu} (k,-q,-p,m_a,m_t,m_a,m_\chi) \nn
    &\qquad\qquad\qquad\qquad\qquad  - D_{\mu\nu} (-k',-q,-p,m_a,m_t,m_a,m_\chi) ] + \mathcal{O}\left(\frac{m_u}{m_t}\right) \bigg\},
\end{align}
which depends on the ALP-quark couplings through the combinations
\begin{align}
    \mC^-_{ut} &= \abs{c^V_{ut}}^2 - \abs{c^A_{ut}}^2 + 2i \Im[c^V_{ut} \qty(c^A_{ut})^*] \gamma^5,\nn
    \mC^+_{ut} &= \abs{c^V_{ut}}^2 + \abs{c^A_{ut}}^2 + 2\Re[c^V_{ut} \qty(c^A_{ut})^*] \,\gamma^5.
\end{align}
The corresponding Feynman diagram is given in~\cref{fig:loop-diagrams}, right. As in the FD case, contributions from $D_{\mu\nu}$ are suppressed, in this case by $|{\bf k^{(')}}|/m_t$.

\bibliographystyle{JHEP_improved}
\bibliography{main}

\end{document}